\documentclass{elsart}
\usepackage{pictex}
\begin{document}
     \newcommand{\p}{\vec{p}\,}
     \newcommand{\q}{\vec{q}\,}

     \newcommand{\tdot}{\!\cdot\!}

     \newcommand{\dd}{{\rm d}}       \newcommand{\ee}{{\rm e}}
     \newcommand{\td}{\!{\rm d}}     \newcommand{\ii}{{\rm i}}
     \newcommand{\re}{{\rm Re}\,}

     \newcommand{\bigeqn}{\begin{eqnarray} && \!\!\! \!\!\! \!\!\!}
     \newcommand{\ie}{i.e.}        \newcommand{\Eq}{Eq.\,}
     \newcommand{\cf}{cf.}         \newcommand{\Eqs}{Eqs.\,}

     \hyphenation{ Min-kow-ski}


%
 \begin{frontmatter}
 \title{The Pole Part of the 1PI Four-Point Function in Light-Cone Gauge 
        Yang-Mills Theory}

 \author[GL]{George Leibbrandt\thanksref{MATH}}, 
 \author[AR]{Al Richardson\thanksref{PHYS}} and
 \author[CP]{C.P. Martin\thanksref{SPAIN}}

 \address[GL]{Department of Mathematics and Statistics, \\
  University of Guelph, Guelph, Ontario, Canada.  N1G 2W1 }
 \address[AR]{Department of Physics, \\
  University of Guelph, Guelph, Ontario, Canada.  N1G 2W1 }
 \address[CP]{Departamento de F\'{\i}sica Te\'orica I, 
         Facultad de Ciencias Fisicas \\
   Universidad Complutense de Madrid,  Madrid-28040, Spain}

 \thanks[MATH]{E-mail: gleibbra@msnet.mathstat.uoguelph.ca}
 \thanks[PHYS]{E-mail: arichard@uoguelph.ca}
 \thanks[SPAIN]{E-mail: carmelo@eucmos.sim.ucm.es}

\begin{abstract}
\begin{center}
{\bf Abstract}
\end{center}
\noindent                      
The complete UV-divergent contribution to the one-loop 1PI four-point 
function of Yang-Mills theory in the light-cone gauge is computed in 
this paper. The formidable UV-divergent contributions arising from
each four-point Feynman diagram yield a succinct final result which 
contains nonlocal terms as expected. These nonlocal contributions 
are consistent with gauge symmetry, and correspond to a nonlocal 
renormalization of the wave function. Renormalization of Yang-Mills 
theory in the light-cone gauge is thus shown explicitly at the 
one-loop level.

\end{abstract}

 \end{frontmatter}
\clearpage
\section{Introduction} Ever since the discovery
\cite{ns:M1,ns:L1,ns:L2,ns:L3} of the $n^{\ast}_{\mu}$-prescription for
the unphysical poles of $(q\tdot n)^{-1}$ in 1982 , the noncovariant
light-cone gauge has gained in usefulness and respect. In fact,
there can be little doubt that the light-cone gauge, defined by $n\tdot
A^a (x)=0,\; n^2 =0$, with $n_{\mu}=(n_0, {\bf n})$ a constant vector,
$n^{\ast}_{\mu}=(n_0,-{\bf n})$, and $n\tdot n^{\ast}=1$, is the most
popular of the axial-type gauges.

The light-cone gauge possesses a number of appealing features: it
picks out the physical degrees of freedom, its prescription allows
a Wick rotation, and the Faddeev-Popov ghosts either decouple from
the associated gauge field, or prove harmless in a perturbative
setting. Moreover, the light-cone gauge often proves more effective
than a covariant gauge (such as the Feynman gauge, or Landau gauge) in
sophisticated models like supersymmetric Yang-Mills and superstrings. By
exploiting the special features of the light-cone gauge, Brink, Lindgren
and Nilsson \cite{brink}, as well as Mandelstam \cite{ns:M1}, succeeded
in demonstrating that the $N=4$ supersymmetric Yang-Mills model is UV
convergent to all orders of perturbation theory, while Green and Schwarz
\cite{supstrng} showed that superstring theory was anomaly-free for the
semi-simple Lie groups $E_8 \times E_8$ and Spin $32/Z_2$.

An important question that was immediately raised after the discovery of
the $n^{\ast}_{\mu}$-prescription, was whether the light-cone gauge was
renormalizable to all orders. In view of the technical idiosyncrasies
of the light-cone gauge Feynman integrals \cite{L:book} generally,
and the appearance of UV divergent {\em nonlocal} expressions in
both the gluon self-energy and the vertex functions, the question
was highly non-trivial. During the ensuing years, the renormalization
structure of Yang-Mills theory in the light-cone gauge was examined by
several researchers, including Bassetto \cite{bas1,bas2}, Bassetto
and his co-workers \cite{basea1,basea2,basea3,basea4}, Lee and Milgram
\cite{leemil1,leemil2}, Andra\v{s}i, Leibbrandt and Nyeo \cite{andrasiea},
Leibbrandt and Nyeo \cite{leinyeo1,leinyeo2}, and Nyeo \cite{ny1,ny2,ny3}.

It is well known that the presence of non-polynomial divergences in the
light-cone gauge greatly complicates the renormalization procedure. The
reason is that these non-polynomial divergences automatically require
non-polynomial counterterms and, at least initially, it appears that one
can construct infinitely many non-polynomial tensor structures that are
consistent with dimensional analysis and the Ward identities. Fortunately,
there exist further constraints on the possible forms of nonlocal
divergences (see Section \ref{sec:disc}). Using these constraints,
Bassetto,Dalbosco and Soldati \cite{Bas:4pt} have predicted the form of
the nonlocal divergent density of $\Gamma$, the generating functional
for 1PI diagrams, from which they derive the nonlocal part of the 1PI
four-point function. Their result allows for renormalization using only
a finite number of both local and nonlocal structures. 
In our explicit calculation of the 1PI
four-point function we obtain not only the predicted {\it nonlocal} tensor 
structure, but also the local part, as well as their relative weights.
It is taken for granted in Bassetto's joint paper that the usual local
part occurs as well.

In 1987, Bassetto, Dalbosco and Soldati gave an inductive proof of the
renormalizability of massless Yang-Mills by showing \cite{Bas:4pt} that
"...despite the appearance of an infinite number of nonlocal divergent
terms, the theory can be made finite to any order in the loop expansion
by introducing a finite number of renormalization constants...", and,
furthermore, that "...the nonlocal structures are completely decoupled
from the physical quantities." The authors were guided in their proof by
previous loop calculations, but there was one 1PI function which had
never been evaluated in the light-cone gauge. This was the infamous 1PI
four-point function, containing box, lynx and fish diagrams, which was
obviously vital for the explicit completion of the renormalization proof
at the one-loop level. Unfortunately, the horrendous algebra inherent in the
computation of the 1PI four-point function was such as to discourage its
evaluation for many years. Bassetto and his co-workers decided to do the
next best thing: by using renormalization arguments, they predicted
that the one-loop 1PI four-point
function possessed a certain tensorial structure in terms of the free
parameters $p_{\mu}$, $n_{\mu}$ and $n^{\ast}_{\mu}$. Their prediction
dealt specifically with the {\em nonlocal} terms in the theory.

The purpose of this paper is to report that the UV divergent
contribution to the  one-loop 1PI four-point function in the light-cone
gauge has now been computed in its entirety, and that the prediction by
Bassetto, Dalbosco and Soldati concerning this function, made over ten
years ago, is in fact correct. Our aim here is to explain and summarize
the computation of this formidable function.

The article is organized thus. The Feynman rules and mathematical tools
are summarized in Sections \ref{sec:fynrules} and \ref{sec:mthtools},
respectively. In Section 4 we evaluate the complete four-point
function consisting, as we shall see, of three specific sets of ``{\em
sub}diagrams'', namely the box diagrams (Section \ref{sec:bxsec}),
lynx diagrams (Section \ref{sec:lxsec}) and fish diagrams (Section
\ref{sec:fshsec}). Only a few of the intermediate results are shown
explicitly to convey to the reader the scope and complexity of the
calculation. Our final expression for the UV divergent contribution to
the 1PI four-point function in the light-cone gauge is presented in Section
\ref{sec:disc}, where we also discuss the role of the nonlocal terms
again, in the context of renormalization. Finally, Appendix A contains
a selection of intermediate expressions for certain box, lynx and fish
diagrams (Eqs. (\ref{eq:bx2lc}), (\ref{eq:lx2lc}) and (\ref{eq:fsh2lc}),
respectively).

\clearpage
\clearpage
\section{Feynman Rules}
\label{sec:fynrules}
In the light-cone gauge,
	\begin{equation}
	n \tdot A^a = 0, \;
	n^2=0 
	\label{eq:lcgf},
	\end{equation}
(the metric signature is $(+, -, -, -)$) the Lagrangian density for $SU(N)$ Yang-Mills theory takes the form 
	\begin{equation} 
	 L= -\frac{1}{4} {F^a}_{\mu \nu} {F_a}^{\mu \nu} - \frac1{2\alpha}{(n\tdot A^a )}^2 
	+ {\overline{\eta}}^a n^{\mu} D^{ab}_{\mu}{\eta}^b ,\; \alpha\rightarrow 0 ,
       \label{eq:lym} 
	\end{equation}
where $F^{a}_{\mu
\nu}={\partial}_{\mu}A^{a}_{\nu}-{\partial}_{\nu}A^{a}_{\mu}+gf^{abc}A^{b}_{\mu}A^{c}_{\nu}$
is the gauge field tensor, $\mu ,\nu = 0,1,2,3$, and $a,b,\ldots = 1,2
...,N^2 -1$.  The covariant derivative is $D^{ab}_{\mu}={\delta}^{ab}
{\partial}_{\mu}+g f^{abc} {A^c}_{\mu}$ and $f^{abc}$ are the group
structure constants. The ghost fields, $\eta$ and $\overline{\eta}$,
appear only in closed loops and obey Fermi statistics.

The Yang-Mills gauge field propagator (Fig.~\ref{f:gaprop}) reads,
for $\alpha=0$,
\begin{equation} \label{eq:prop}
G^{ab}_{\mu \nu}(q) = \frac{-i {\delta}^{ab}}{q^2 +i \epsilon}
\bigg[ g_{\mu \nu} -\frac{q_\mu n_\nu+q_\nu n_\mu}{q \tdot n} \bigg] ,\; \epsilon > 0.
\end{equation}
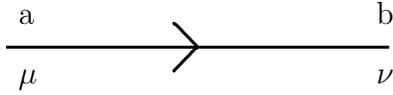
\begin{figure}[h,b,t]
\setlength{\unitlength}{0.012500in}%
\begin{center}
\begin{picture}(160,34)(80,745)
\thicklines
\put( 80,760){\line( 1, 0){160}}
\multiput(160,760)(-0.40000,0.40000){26}{\makebox(0.4444,0.6667){.}}
\multiput(160,760)(-0.40000,-0.40000){26}{\makebox(0.4444,0.6667){.}}
\put( 85,770){\makebox(0,0)[lb]{\smash{a}}}
\put(235,770){\makebox(0,0)[lb]{\smash{b}}}
\put( 85,745){\makebox(0,0)[lb]{\smash{$\mu$}}}
\put(235,745){\makebox(0,0)[lb]{\smash{$\nu$}}}
\end{picture}
\end{center}
\vskip .2in
\caption{ \em Light-Cone Gauge Yang-Mills Propagator}
\label{f:gaprop}
\vskip .2in
\end{figure}

The ghost-ghost-gluon vertex is given by
\begin{equation}  \label{eq:gggvtex}
U^{abc}_{\mu}(p,q,r)  =  - i g f^{abc} n_{\mu} {\delta}^{4} (q+p-r),
\end{equation}
so that any interaction between ghost and gluon involves the contraction
of the gluon propagator with $n^{\mu}$. Thus,
\begin{equation}
n^{\mu}G^{ab}_{\mu \nu}(q)= \frac{-i{\delta}^{ab}}{q^2+i\epsilon}n^{\mu} 
\bigg[ g_{\mu \nu} -\frac{q_{\mu} n_{\nu}+q_{\nu}n_{\mu}}{q \tdot n} \bigg]
= n_{\nu} - n_{\nu} -q_{\nu}\frac{n^2}{q \tdot n} = 0, 
\end{equation}
so that any ghost-gluon interaction term vanishes, since $n_{\mu}$ is
light-like. Although the ghost propagator is not needed for the present
calculations, we list it here for completeness:
\begin{equation}
G^{ab}=\frac{-i {\delta}^{ab}}{q\tdot n}.
\end{equation}

The bare three-gluon vertex (Fig.~\ref{f:fvtex3}) reads
\begin{equation}
V^{abc}_{\mu \nu \rho}(p,q,r)  =  - g f^{abc} \big[ (q - r)_{\mu}
g_{\nu \rho} + (r-p)_{\nu} g_{\mu \rho} +(p-q)_{\rho} g_{\mu \nu} \big],
\end{equation}
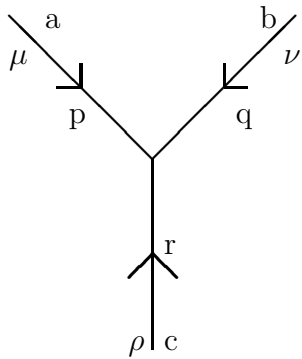
\begin{figure}[h,b,t]
\vskip .1in
\setlength{\unitlength}{0.012500in}%
\begin{center}
\begin{picture}(120,144)(100,600)
\thicklines
\put(160,680){\line( 0,-1){ 80}}
\put(160,680){\line(-1, 1){ 60}}
\put(130,710){\line( 0, 1){ 10}}
\put(130,710){\line(-1, 0){ 10}}
\multiput(160,640)(0.40000,-0.40000){26}{\makebox(0.4444,0.6667){.}}
\multiput(160,640)(-0.40000,-0.40000){26}{\makebox(0.4444,0.6667){.}}
\put(160,680){\line( 1, 1){ 60}}
\put(190,710){\line( 1, 0){ 10}}
\put(190,720){\line( 0,-1){ 10}}
\put(115,735){\makebox(0,0)[lb]{\smash{a}}}
\put(205,735){\makebox(0,0)[lb]{\smash{b}}}
\put(195,695){\makebox(0,0)[lb]{\smash{q}}}
\put(125,695){\makebox(0,0)[lb]{\smash{p}}}
\put(165,640){\makebox(0,0)[lb]{\smash{r}}}
\put(150,600){\makebox(0,0)[lb]{\smash{$\rho$}}}
\put(165,600){\makebox(0,0)[lb]{\smash{c}}}
\put(100,720){\makebox(0,0)[lb]{\smash{$\mu$}}}
\put(215,720){\makebox(0,0)[lb]{\smash{$\nu$}}}
\end{picture}
\end{center}
\caption{ \em Three-Gluon Vertex}
\label{f:fvtex3}
\vskip .1in
\end{figure}
and the bare four-gluon vertex (Fig.~\ref{f:fvtex4}) is
\begin{eqnarray} \label{eq:fgvtex}
\hspace{.75in}V^{abcd}_{\mu \nu \rho \sigma}  =  - i g^2 &[& 
f_{abe} f_{cde} (g_{\mu \rho} g_{\nu \sigma} - g_{\mu \sigma} g_{\nu \rho} ) \\
&+& f_{ace} f_{dbe} ( g_{\mu \sigma} g_{\nu \rho} - g_{\mu \nu} g_{\rho \sigma} )\nonumber \\
&+& f_{ade} f_{bce} ( g_{\mu \nu} g_{\rho \sigma} - g_{\mu \rho} g_{\nu \sigma} ) ]. \nonumber
\end{eqnarray}
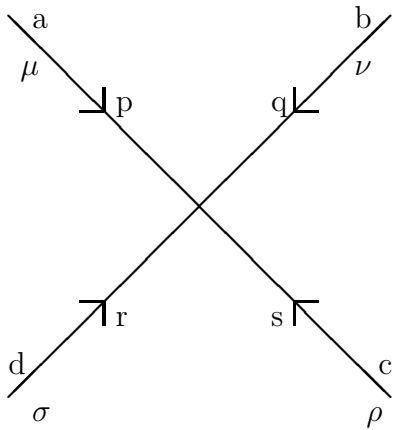
\begin{figure}[h]
\vskip .1in
\setlength{\unitlength}{0.012500in}%
\begin{center}
\begin{picture}(160,174)(80,590)
\thicklines
\put(160,680){\line(-1, 1){ 80}}
\put(120,720){\line( 0, 1){ 10}}
\put(120,720){\line(-1, 0){ 10}}
\put(160,680){\line( 1, 1){ 80}}
\put(160,680){\line( 1,-1){ 80}}
\put(160,680){\line(-1,-1){ 80}}
\put(200,720){\line( 0, 1){ 10}}
\put(210,720){\line(-1, 0){ 10}}
\put(210,640){\line(-1, 0){ 10}}
\put(120,630){\line( 0, 1){ 10}}
\put(120,640){\line(-1, 0){ 10}}
\put(200,630){\line( 0, 1){ 10}}
\put( 90,755){\makebox(0,0)[lb]{\smash{a}}}
\put(225,755){\makebox(0,0)[lb]{\smash{b}}}
\put(225,735){\makebox(0,0)[lb]{\smash{$\nu$}}}
\put( 85,735){\makebox(0,0)[lb]{\smash{$\mu$}}}
\put(125,720){\makebox(0,0)[lb]{\smash{p}}}
\put(190,720){\makebox(0,0)[lb]{\smash{q}}}
\put(125,630){\makebox(0,0)[lb]{\smash{r}}}
\put(190,630){\makebox(0,0)[lb]{\smash{s}}}
\put(235,610){\makebox(0,0)[lb]{\smash{c}}}
\put( 80,610){\makebox(0,0)[lb]{\smash{d}}}
\put( 90,590){\makebox(0,0)[lb]{\smash{$\sigma$}}}
\put(230,590){\makebox(0,0)[lb]{\smash{$\rho$}}}
\end{picture}
\end{center}
\caption{ \em Four-Gluon Vertex}
\label{f:fvtex4}
\vskip .1in
\end{figure}

\clearpage
\clearpage
\section{Mathematical Tools}\label{sec:mthtools}
In this section, we shall discuss the technical aspects relevant to the calculation of the four-point function.

We work in the framework of dimensional regularization, where $2\omega$ defines the dimensionality of complex
 space-time, and recall that massless tadpoles, such as
\begin{equation}
\int d^{2\omega}q \frac{1}{q^2}, \;\int d^{2\omega}q \frac{1}{q\tdot n},\;\int d^{2\omega}q \frac{1}{q^2 (q\tdot n)}, etc.
\end{equation}
are set to zero. 

The spurious poles of $(q\tdot n)^{-1}$ are treated with the $n^{\ast}_{\mu}$-prescription \cite{ns:M1,ns:L1,ns:L2,ns:L3},
\begin{equation}\label{eq:nspre}
\frac{1}{q\tdot n} = lim_{\epsilon \longrightarrow 0} \frac{q \tdot n^{\ast}}{(q\tdot n)( q\tdot n^{\ast})+i\epsilon },\;\epsilon >0,
\end{equation}
which allows a Wick rotation and is consistent with power counting. 
Note that care must be taken when using power counting arguments, since the degree of divergence $\alpha$ differs for $q_{\perp}$ 
and $q_{\parallel}$, where
\begin{eqnarray}
& & \hspace{1.2in} q\tdot n  =  q_{\parallel}\tdot n, \\
& & \hspace{1.2in} q_{\perp}\tdot n  =  0. \nonumber
\end{eqnarray} 
For example, the gauge propagator (\ref{eq:prop}) has a degree of
divergence $\alpha = -2$, when $q \rightarrow \infty$ along any
direction. However, with respect to large $q_{\perp}$ behaviour, the
covariant part of the propagator has degree of divergence
${\alpha}_{\perp}=-2$, while the noncovariant parts carry degree
${\alpha}_{\perp}=-1$.  When encountering multiple $(q \cdot n)^{-1}$
factors coming from the gauge propagator (\ref{eq:prop}), we shall use
the separation formula
\begin{equation}\label{eq:sepn}
\frac{1}{q \cdot n (q - p) \cdot n} = \frac{1}{p \cdot n}
\left[ \frac{1}{(q-p) \cdot n} - \frac{1}{q \cdot n} \right] , p \cdot n \not = 0,
\end{equation}
which helps unveil the nonlocal structure of the UV divergence of the
Green functions \cite{gl3}.

The
separation formula (\ref{eq:sepn}) is applied as many times as is
necessary to yield integrals containing only a single noncovariant
denominator; a shift in the integration variable $k$, $k=q-p$, is used
to reduce as many integrals as possible to tadpole integrals, which
are known to vanish in dimensional regularization.

To simplify complicated Feynman integrals, we split the gauge propagator
(\ref{eq:prop}) into light-cone (LC) and non-light-cone (NLC) propagators,
$G^{a b}_{\mu \nu} = {\delta}^{ab} (G^{LC}_{\mu \nu} + G^{NLC}_{\mu \nu}
)$, where $G^{LC}_{\mu \nu}$ contains the noncovariant vector $n_{\mu}$,
while $G^{NLC}_{\mu \nu}$ contains only covariant parts:
\begin{eqnarray}	\label{eq:lcprop}
\hspace{1.2in} & &G^{a b}_{\mu \nu} = {\delta}^{ab} (G^{LC}_{\mu \nu} + G^{NLC}_{\mu \nu} ),\\ 
& & G^{LC}_{\mu \nu}(q) =  \frac{i}{q^{2} + i \epsilon} 
\bigg( \frac{ q_{\mu}n_{\nu}+q_{\nu}n_{\mu} }{q \tdot n} \bigg), \nonumber \\
& & G^{NLC}_{\mu \nu}(q) = \,- \frac{i}{q^{2}+ i \epsilon} g_{\mu \nu}. \nonumber
\end{eqnarray}

Finally, in order to handle the vast number of terms in the intermediate
stages of the calculation, we employed the program MAPLE, with the add-on
package HIP-MAPLE, which allowed us to manage simple operations on large
expressions, such as permutations of indices, additions, etc. The main
steps involved in the computation of the four-point function may be
summarized as follows.  \begin{enumerate}
\item The Feynman integrals corresponding to the box diagram
(Fig.~\ref{fig:box0}), lynx diagram (Fig.~\ref{fig:lynx0}) and fish
diagram (Fig.~\ref{fig:fish0}) are written down, and the propagators split
according to Eqs. (\ref{eq:lcprop}), generating 28 major ``subdiagrams''.

\item After simplification, each integral in the 28 ``subdiagrams'' is
identified as belonging to one of 44 classes of typical integrals,
$[I^{\{\mu\}}_{\{ a\}}(\{p\} ){\big]}_i$, where $\{ \mu \}$ indicates the
relevant set of Lorentz indices, $\{ a \}$ the set of colour indices, $\{
p \}$ the set of (arbitrary) external momenta and where $i=1,\ldots, 44$.

\item The UV divergent bit of each of the 44 classes of integrals
is evaluated by hand, for general external momenta. The results are
fed into the symbolic manipulation program MAPLE, with the add-on
package HIP-MAPLE, which greatly simplifies algebraic manipulations of
four-vectors. Unfortunately, HIP-MAPLE is unsupported and will not work,
without significant revisions, with MAPLE version three or higher.

\item The integrated results of each ``subdiagram'' are then entered into the
computer. The ability of MAPLE to easily substitute specific sets of
external momenta and indices into the various
integrals is invaluable at this stage.

\item The complete expression for the UV divergent bit of 1PI the four-point
function is obtained by summing the box, lynx and fish diagrams of
Section~\ref{sec:fptfn}, along with their `partner' diagrams which are generated
by taking all topologically distinct permutations of external indices
and momenta.  The integrated results for all ``subdiagrams'' and `partners'
are combined and simplified, yielding the concise final result in
Eq.~(\ref{eq:result}).  
\end{enumerate}

\clearpage
\clearpage
\section{The 1PI Four-Point Function}
\label{sec:fptfn}
The 1PI four-point function in light-cone Yang-Mills theory is comprised of
three diagrams: the box diagram (Fig.~\ref{fig:box0}), the lynx diagram
(Fig.~\ref{fig:lynx0}) and the fish diagram (Fig.~\ref{fig:fish0}). We
shall deal with each diagram separately. 

To obtain the complete 1PI four-point function, one must include not only
the representative diagrams (discussed in this section), but also the diagrams
generated by taking all topologically distinct permutations of external
indices and momenta. Once the calculations in Sections \ref{sec:bxsec},
\ref{sec:lxsec} and \ref{sec:fshsec} have been completed, the final
permutations may be readily generated using MAPLE. Accordingly, we shall
only present one specific ordering of the external indices.

Before turning to the explicit calculations, it may be helpful to comment
briefly on the possible nonlocal structures allowed in the 1PI four-point
function. Each noncovariant propagator $G^{LC}$ contributes a factor
$[(q-p)\tdot n]^{-1}$. Since the divergent parts of basic one-loop
integrals (i.e. those with only a single factor of $[(q-p)\tdot n]^{-1}$)
are {\em local} functions of the external momenta \cite{l:nloc}, the
pole part of an integral with only one noncovariant propagator is local,
and we use the separation formula Eq. (\ref{eq:sepn}) as many times as
necessary to reduce an integral with multiple noncovariant propagators to
a sum of basic one-loop integrals. Furthermore, each application of the
separation formula introduces one nonlocal factor $(p_i \tdot n)^{-1}$
in the external momenta $p_i$, and these are the only non-polynomial
factors. Dimensional analysis of the fish, lynx and box integrals
Eqs. (\ref{eq:boxeqn}), (\ref{eq:lynxeqn}) and (\ref{eq:fisheqn}), reveals
that the 1PI four-point function has units of $[p]^0 [n]^0 [n^{\ast}]^0$,
so we can tell that, for the fish diagram, the integrated expression
can contain only factors of the form \begin{equation}
f^{u_1 u_2 u_3 u_4}_{0}, \; \frac{f^{u_1 u_2 u_3 u_4}_{1}}{p\tdot n},\frac{f^{u_1 u_2 u_3 u_4}_{2}}{(p\tdot n)(k\tdot n)},
\end{equation}
where the subscripts $0,1,2$ indicate the power of momentum in
the expression $f$; the poles of $(p\tdot n)^{-1}$ are treated according to the
$n^{\ast}_{\mu}$-prescription and colour indices have been omitted. In the
case of the lynx diagram, an additional factor of the form $$\frac{f^{u_1
u_2 u_3 u_4}_{3}}{(p\tdot n)(k\tdot n)(q\tdot n)}$$ can occur, along with
those listed above, while in the box diagram a fourth factor, $$\frac{f^{u_1
u_2 u_3 u_4}_{4}}{(p\tdot n)(k\tdot n)(q\tdot n)(r\tdot n)},$$ may
also appear.

\subsection{Box Diagrams}
\label{sec:bxsec}
The complete pole part of the box diagram, shown in Fig.~\ref{fig:box0}
is difficult to calculate by virtue of the sheer number of Feynman
integrals. It leads to the following expression:
\begin{eqnarray}
\label{eq:boxeqn}
\hspace{.5in} J^{box}=\bigg{\{} & & \int \frac{d^{2 \omega}q}{(2\pi )^{2\omega}}
{\delta}^{ab}G_{\mu \nu}(q)V^{u_1 \mu \tau}_{a_1 a h}(p_1 ,q,q+p_4 ) \\
& & {\delta}^{cd}G_{\alpha \beta}(q+p_2 )V^{u_2 \alpha \nu}_{a_2 c b}(p_2 ,-q-p_2 ,q) \nonumber \\
& & {\delta}^{ef}G_{\rho \sigma}(q+p_2 +p_3 )\nonumber \\
& & V^{u_3 \rho \beta}_{a_3 e d}(p_3 ,-q-p_2 -p_3 ,q+p_2 )\nonumber \\
& & {\delta}^{gh}G_{\eta \tau}(q +p_2 +p_3 +p_4 )\nonumber \\
& & V^{u_4 \eta \sigma}_{a_4 g f}(p_4 ,-q-p_2 -p_3 -p_4 ,q+p_2 +p_3 ) \bigg{\} }\nonumber \\
& & \delta^{2\omega}(p_1 +p_2 +p_3 +p_4 ) \nonumber .
\end{eqnarray}
\begin{figure}[h]
\begin{center}
\input{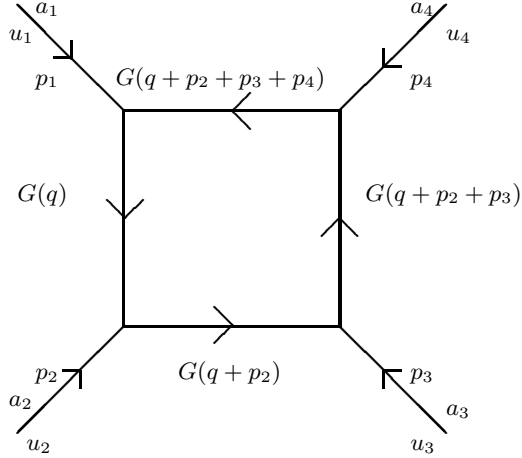}
\end{center}
\caption{ \em Complete Box Diagram}
\label{fig:box0}
\end{figure}

With the decomposition given in Eq.~(\ref{eq:lcprop}), the complete box diagram
breaks down into 16 ``subdiagrams'', each having only one factor of $G^{LC}$,
or $G^{NLC}$, along any internal line. Several of the ``subdiagrams'' are
related to each other by permutations of the triplet of external indices
and momenta, ($a_i$, $u_i$, $p_i$), $i=1,2,3,4$ (see Eq. (\ref{eq:4p1lc})).

The simplest ``subdiagram'' (Fig.~\ref{fig:box0lc}) contains {\em no}
noncovariant terms $G^{LC}$, which severely restricts the possible
tensor structures appearing in the integrated result.  The integral
corresponding to this subdiagram has the form
\begin{eqnarray}
\label{eq:box0lceqn}
\hspace{.5in} J^{box}_{0}= \bigg{\{ } & & \int \frac{d^{2 \omega}q}{(2\pi )^{2\omega}}
{\delta}^{ab}G^{NLC}_{\mu \nu}(q)V^{u_1 \mu \tau}_{a_1 a h}(p_1 ,q,q+p_4 ) \\ & & {\delta}^{cd}G^{NLC}_{\alpha \beta}(q+p_2)V^{u_2 \alpha \nu}_{a_2 c b}(p_2 ,-q-p_2 ,q) \nonumber \\
& & {\delta}^{ef}G^{NLC}_{\rho \sigma}(q+p_2 +p_3 )\nonumber \\
& & V^{u_3 \rho \beta}_{a_3 e d}(p_3 ,-q-p_2 -p_3 ,q+p_2 ) \nonumber\\
& & {\delta}^{gh}G^{NLC}_{\eta \tau}(q +p_2 +p_3 +p_4 )\nonumber \\
& & V^{u_4 \eta \sigma}_{a_4 g f}(p_4 ,-q-p_2 -p_3 -p_4 ,q+p_2+p_3 )\bigg{\}}\nonumber \\
& & \delta^{2\omega}(p_1 +p_2 +p_3 +p_4 );\nonumber
\end{eqnarray}
\begin{figure}[h]
\begin{center}
\input{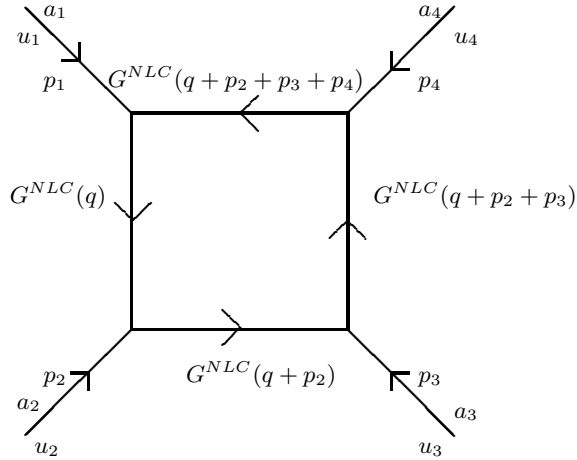}
\end{center}
\caption{ \em Box Diagram: No Noncovariant Propagators }
\label{fig:box0lc}
\end{figure}
and yields the simple expression
\begin{eqnarray}
J^{box}_{0} =\frac{\overline{I} \, f^{a_1 ab}f^{a_2 bc}f^{a_3 cd}f^{a_4 da}}{12}
\,\bigg{\{} 47\,\big[ g_{u_3 u_4}\, g_{u_1 u_2} &+& g_{u_2 u_3}\, g_{u_1 u_4}\big]\\ 
+ 17\, g_{u_1 u_3}g_{u_2 u_4}& &\bigg{\}}\nonumber ,
\end{eqnarray}
where $\overline{I}= i {\pi}^2 / (2-\omega )$ . 

For ``subdiagrams'' with one or more $G^{LC}$ factors, the
integrated expressions become increasingly complicated, often containing
hundreds of individual terms. To see this, consider first the set of
diagrams (Fig.~\ref{fig:box1lc}) including precisely {\it one} noncovariant
propagator $G^{LC}_{\mu \nu}$, which leads to the formula
\begin{eqnarray}
\label{eq:4p1lc}
J^{box}_{1}= \bigg{\{ } & & \int  \frac{d^{2 \omega}q}{(2\pi)^{2\omega}}{\delta}^{ab}{\delta}^{cd}{\delta}^{ef}{\delta}^{gh}
G^{LC}_{\mu \nu}(q)V^{u_1 \mu \tau}_{a_1 a h}(p_1 ,q,q+p_4 ) \\
& & G^{NLC}_{\alpha \beta}(q+p_2)V^{u_2 \alpha \nu}_{a_2 c b}(p_2 ,-q-p_2 ,q) \nonumber \\
& & G^{NLC}_{\rho \sigma}(q+p_2 +p_3 )\nonumber \\
& & V^{u_3 \rho \beta}_{a_3 e d}(p_3 ,-q-p_2 -p_3 ,q+p_2 )\nonumber \\
& & G^{NLC}_{\eta \tau}(q +p_2 +p_3 +p_4)\nonumber \\
& & V^{u_4 \eta \sigma}_{a_4 g f}(p_4 ,-q-p_2 -p_3 -p_4 ,q+p_2+p_3 ) \nonumber \\
&+&(1\rightarrow 2,2\rightarrow 3,3\rightarrow 4, 4\rightarrow 1)\nonumber \\
&+&(1\rightarrow 3,2\rightarrow 4,3\rightarrow 1, 4\rightarrow 2)\nonumber \\
&+&(1\rightarrow 4,2\rightarrow 1,3\rightarrow 2, 4\rightarrow 3)\bigg{\}}
\delta^{2\omega}(p_1 +p_2 +p_3 +p_4 );\nonumber
\end{eqnarray}
\begin{figure}[h]
\begin{center}
\input{box1lc-a.te}
\vskip .1in
\input{box1lc-b.te}
\end{center}
\caption{ \em Box Diagrams: One Noncovariant Propagator}
\label{fig:box1lc}
\end{figure} 
here, ($1\rightarrow 2$) indicates the exchange of both external indices
and momenta; for example: $a_1 \rightarrow a_2 ,u_1 \rightarrow u_2 ,p_1
\rightarrow p_2 $, and so on. Although all four of these ``subdiagrams''
are related by symmetry transformations, the possibility of errors in
the calculation makes it advisable to compute explicitly at least two
of the ``subdiagrams''. Evaluation of the integrals in Eq.~(\ref{eq:4p1lc})
leads to several hundred terms, but fortuitous cancellation produces
the manageable expression 
\begin{eqnarray}
\hspace{.5in} J^{box}_{1}=\frac{\overline{I}\, f^{a_1 ab}f^{a_2 bc} f^{a_3 cd}f^{a_4 da}}{  n \tdot n^{\ast}} & &\\
\bigg{\{ }
 - 4 \big[
\;n_{ u_1 }\, n^{\ast}_{ u_4 }\, n^{\ast}_{ u_3 }\, n_{ u_2 }
& & \mbox{} + n_{ u_3 }\, n^{\ast}_{ u_1 }\, n^{\ast}_{ u_4 }\, n_{ u_2 }\nonumber \\
+ n_{ u_1 }\, n^{\ast}_{ u_4 }\, n^{\ast}_{ u_2 }\, n_{ u_3 }
& & \mbox{} + n_{ u_4 }\, n^{\ast}_{ u_1 }\, n^{\ast}_{ u_2 }\, n_{ u_3 }\nonumber \\
+ n_{ u_4 }\, n^{\ast}_{ u_1 }\, n^{\ast}_{ u_3 }\, n_{ u_2 }
& & \mbox + n_{ u_4 }\, n^{\ast}_{ u_2 }\, n^{\ast}_{ u_3 }\, n_{ u_1 }\, \big] \nonumber \\
+ 4 (n\tdot n^{\ast}) \big[\;n_{ u_4 }\, n^{\ast}_{ u_3 }\> g_{ u_1 u_2 }
& & \mbox{} + n_{ u_1 }\, n^{\ast}_{ u_4 }\> g_{ u_2 u_3 } \nonumber \\
+ n_{ u_1 }\, n^{\ast}_{ u_2 }\> g_{ u_3 u_4 } 
& & \mbox + n_{ u_4}\, n^{\ast}_{ u_1}\> g_{ u_2 u_3} \nonumber \\
+ n_{ u_3}\, n^{\ast}_{ u_4}\> g_{ u_1 u_2}
& & \mbox{} + n_{ u_2}\, n^{\ast}_{ u_1}\> g_{ u_3 u_4} \nonumber \\
+ n_{ u_2}\, n^{\ast}_{ u_3}\> g_{ u_1 u_4}
& & \mbox{} + n_{ u_3}\, n^{\ast}_{ u_2}\> g_{ u_1 u_4} \big] \nonumber \\
+ 6(n\tdot n^{\ast}) \big[\;n_{ u_1}\, n^{\ast}_{ u_3}\> g_{ u_2 u_4}
& & \mbox{} + n_{ u_2}\, n^{\ast}_{ u_4}\> g_{ u_1 u_3}\nonumber \\
+ n_{ u_4}\, n^{\ast}_{ u_2}\> g_{ u_1 u_3}
& & \mbox{} + n_{ u_3}\, n^{\ast}_{ u_1}\> g_{ u_2 u_4} \big] \nonumber \\
- \frac {25}{3}(n\tdot n^{\ast})^2 \big[
\; g_{ u_1 u_2}\> g_{ u_3 u_4} 
& & \mbox{} +  g_{ u_2 u_3}\> g_{ u_1 u_4}\big]\nonumber \\
- \frac {13}{3}(n\tdot n^{\ast})^2 \big[
\; g_{ u_1 u_3}\> g_{ u_2 u_4}\big] & & \bigg{\} }. \nonumber
\end{eqnarray}
With the introduction of the single $G^{LC}(q)$ factor, we already
begin to see the emergence of noncovariant terms. Of course, there
are no nonlocal terms, since the separation formula Eq.~(\ref{eq:sepn})
has not been required as yet.

The set of ``subdiagrams'' containing {\em two} $G^{LC}$ factors
(Fig.~\ref{fig:box2lc}) is represented by $J^{box}_2$, where
\begin{eqnarray}
\label{eq:4p2lc}
J^{box}_{2}=
\bigg{\{ } & & \int  \frac{d^{2 \omega}q}{(2\pi)^{2\omega}}{\delta}^{ab}{\delta}^{cd}{\delta}^{ef}{\delta}^{gh}
G^{LC}_{\mu \nu}(q)\\
& & V^{u_1 \mu \tau}_{a_1 a h}(p_1 ,q,q+p_4 )\nonumber \\
& & G^{LC}_{\alpha \beta}(q+p_2)V^{u_2 \alpha \nu}_{a_2 c b}(p_2 ,-q-p_2 ,q) \nonumber \\
& & G^{NLC}_{\rho \sigma}(q+p_2 +p_3 )\nonumber \\
& & V^{u_3 \rho \beta}_{a_3 e d}(p_3 ,-q-p_2 -p_3 ,q+p_2 )\nonumber \\
& & G^{NLC}_{\eta \tau}(q +p_2 +p_3 +p_4 )\nonumber \\
& & V^{u_4 \eta \sigma}_{a_4 g f}(p_4 ,-q-p_2 -p_3 -p_4 ,q+p_2+p_3 ) \nonumber \\
&+&(1\rightarrow 2,2\rightarrow 3,3\rightarrow 4, 4\rightarrow 1)\nonumber \\
&+&(1\rightarrow 3,2\rightarrow 4,3\rightarrow 1, 4\rightarrow 2)\nonumber \\
&+&(1\rightarrow 4,2\rightarrow 1,3\rightarrow 2, 4\rightarrow 3)
\bigg{\}}\delta^{2\omega}(p_1 +p_2 +p_3 +p_4 )\nonumber \\
+ \bigg{\{ } & & \int \frac{d^{2 \omega}q}{(2\pi)^{2\omega}}{\delta}^{ab}{\delta}^{cd}{\delta}^{ef}{\delta}^{gh}
G^{LC}_{\mu \nu}(q)V^{u_1 \mu \tau}_{a_1 a h}(p_1 ,q,q+p_4 ) \nonumber \\
& & G^{NLC}_{\alpha \beta}(q+p_2)V^{u_2 \alpha \nu}_{a_2 c b}(p_2 ,-q-p_2 ,q) \nonumber \\
& & G^{LC}_{\rho \sigma}(q+p_2 +p_3 )\nonumber \\
& & V^{u_3 \rho \beta}_{a_3 e d}(p_3 ,-q-p_2 -p_3 ,q+p_2 ) \nonumber \\
& & G^{NLC}_{\eta \tau}(q +p_2 +p_3 +p_4)\nonumber \\
& & V^{u_4 \eta \sigma}_{a_4 g f}(p_4 ,-q-p_2 -p_3 -p_4 ,q+p_2+p_3 ) \nonumber \\
&+&(1\rightarrow 2,2\rightarrow 3,3\rightarrow 4, 4\rightarrow 1)
\bigg{\}} \delta^{2\omega}(p_1 +p_2 +p_3 +p_4 ).\nonumber
\end{eqnarray}
\begin{figure}[bh]
\begin{center}
\input{box2lc-a.te}
\vskip .1in
\input{box2lc-b.te}
\vskip .1in
\input{box2lc-c.te}
\end{center}
\caption{ \em Box Diagrams: Two Noncovariant Propagators}
\label{fig:box2lc}
\end{figure}
Evaluation of Eq.~(\ref{eq:4p2lc}) yields the expression (\ref{eq:bx2lc}) in Appendix A.

The next contribution to the complete box comes from ``subdiagrams''
containing three noncovariant propagator factors (Fig.~\ref{fig:box3lc}),
and has the unintegrated form \begin{eqnarray}
\label{eq:4p3lc}
J^{box}_{3}= \bigg{\{ } & & \int \frac{d^{2 \omega}q}{(2\pi)^{2\omega}}{\delta}^{ab}{\delta}^{cd}{\delta}^{ef}{\delta}^{gh}
G^{LC}_{\mu \nu}(q)V^{u_1 \mu \tau}_{a_1 a h}(p_1 ,q,q+p_4 ) \\
& & G^{LC}_{\alpha \beta}(q+p_2)V^{u_2 \alpha \nu}_{a_2 c b}(p_2 ,-q-p_2 ,q) \nonumber \\
& & G^{LC}_{\rho \sigma}(q+p_2 +p_3 )V^{u_3 \rho \beta}_{a_3 e d}(p_3 ,-q-p_2 -p_3 ,q+p_2 ) \nonumber \\
& & G^{NLC}_{\eta \tau}(q +p_2 +p_3 +p_4 )\nonumber \\
& & V^{u_4 \eta \sigma}_{a_4 g f}(p_4 ,-q-p_2 -p_3 -p_4 ,q+p_2+p_3 ) \nonumber \\
&+&(1\rightarrow 2,2\rightarrow 3,3\rightarrow 4, 4\rightarrow 1)\nonumber \\
&+&(1\rightarrow 3,2\rightarrow 4,3\rightarrow 1, 4\rightarrow 2)\nonumber \\
&+&(1\rightarrow 4,2\rightarrow 1,3\rightarrow 2, 4\rightarrow 3)
\bigg{\} }\delta^{2\omega}(p_1 +p_2 +p_3 +p_4 ).\nonumber 
\end{eqnarray}
\begin{figure}[h]
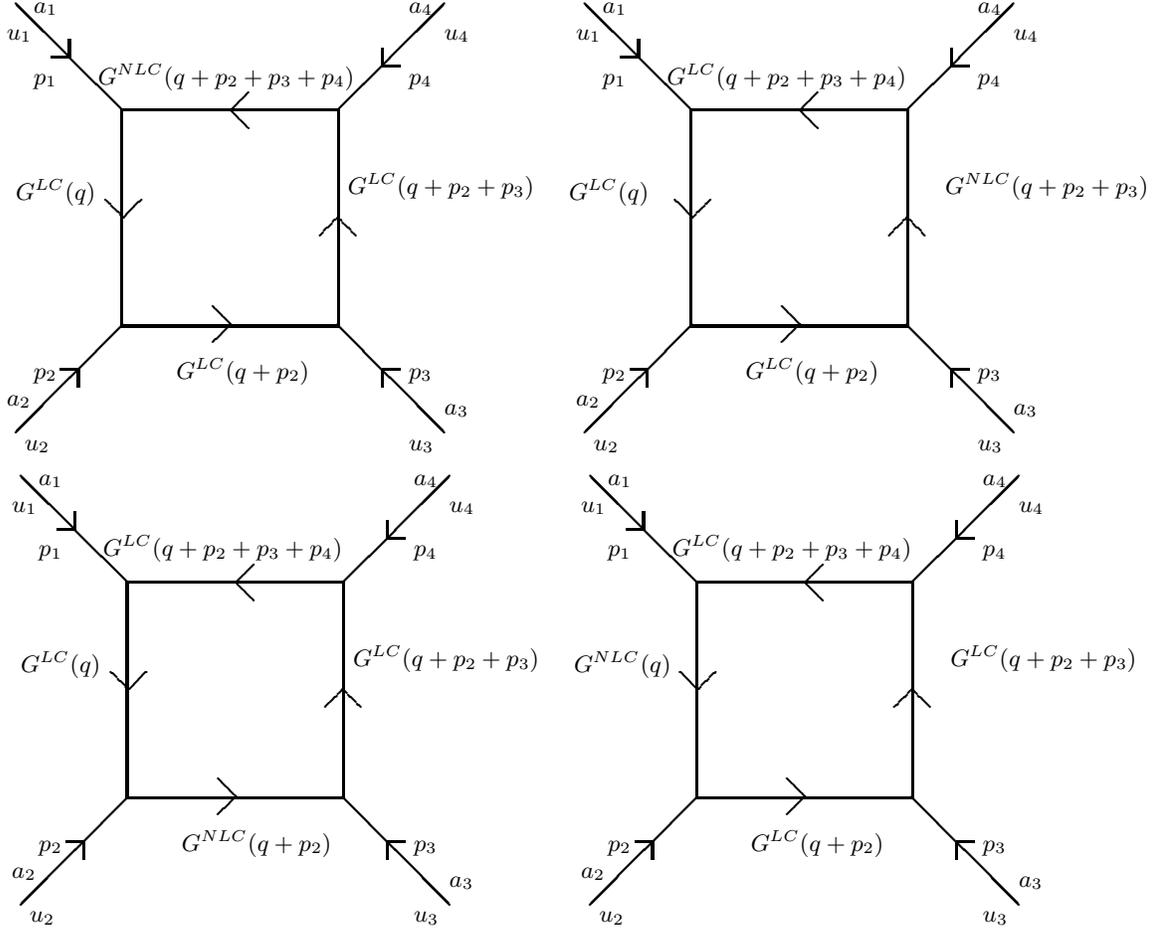

\begin{center}
\input{box3lc-a.te}
\vskip .1in
\input{box3lc-b.te}
\end{center}
\caption{ \em Box Diagrams: Three Noncovariant Propagators}
\label{fig:box3lc}
\end{figure}
We shall make no attempt to state the pole part of the integrated result
of Eq.~(\ref{eq:4p3lc}), consisting as it does of some $653$
terms. Fortunately, the UV contribution to $J^{box}_{3}$ is cancelled in its entirety by the
corresponding terms, both in its symmetry partners, and in $J^{lynx}_{3}$
(see below). In fact, it will turn out that this entire pole part of the box diagram is
cancelled in the complete 1PI one-loop four-point function.

The last contribution to the complete box comes from the diagram with four
internal noncovariant propagators $G^{LC}$ (Fig.~\ref{fig:box4lc});
the associated integral reads
\begin{eqnarray}
\label{eq:4p4lc}
J^{box}_{4}= \bigg{\{ } & & \int  \frac{d^{2 \omega}q}{(2\pi)^{2\omega}}{\delta}^{ab}{\delta}^{cd}{\delta}^{ef}{\delta}^{gh}
G^{LC}_{\mu \nu}(q)V^{u_1 \mu \tau}_{a_1 a h}(p_1 ,q,q+p_4 ) \\
& & G^{LC}_{\alpha \beta}(q+p_2)V^{u_2 \alpha \nu}_{a_2 c b}(p_2 ,-q-p_2 ,q) \nonumber \\
& & G^{LC}_{\rho \sigma}(q+p_2 +p_3 )\nonumber \\
& & V^{u_3 \rho \beta}_{a_3 e d}(p_3 ,-q-p_2 -p_3 ,q+p_2 ) \nonumber \\
& & G^{LC}_{\eta \tau}(q +p_2 +p_3 +p_4 )\nonumber \\
& & V^{u_4 \eta \sigma}_{a_4 g f}(p_4 ,-q-p_2 -p_3 -p_4 ,q+p_2+p_3 )\bigg{\}}\nonumber\\
& & \delta^{2\omega}(p_1 +p_2 +p_3 +p_4 ).\nonumber
\end{eqnarray}
\begin{figure}[h]
\begin{center}
\input{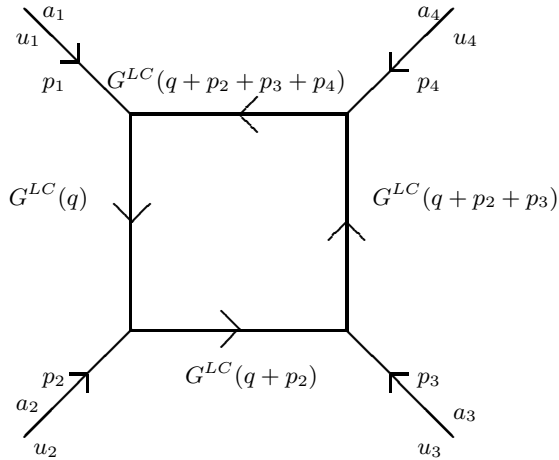}
\end{center}
\caption{ \em Box Diagram: Four Noncovariant Propagators}
\label{fig:box4lc}
\end{figure}
Once again, the intermediate answer from the integral (\ref{eq:4p4lc})
is too unwieldy ($611$ individual terms) and will not be shown explicitly.

\subsection{Lynx Diagrams} \label{sec:lxsec}
In this section, we shall discuss the contributions from the complete lynx
diagram, Fig.~\ref{fig:lynx0}, quoting several intermediate expressions. We
notice that, in view of the structure of the bare four-gluon vertex
(\ref{eq:fgvtex}), all lynx ``subdiagrams'' will be symmetric with respect
to the following interchange: $a_1\leftrightarrow a_4 ,u_1\leftrightarrow
u_4 ,p_1\leftrightarrow p_4 $. Thus,
\begin{eqnarray}
\label{eq:lynxeqn}
J^{lynx}= \bigg{\{ } & & \int \frac{d^{2 \omega}q}{(2\pi)^{2\omega}}{\delta}^{ab}{\delta}^{cd}{\delta}^{ef}\\
& & G_{\mu \nu}(q)V^{u_2 \nu \alpha}_{a_2 b c}(p_2 ,q, -q-p_2 ) \nonumber \\
& & G_{\alpha \beta}(q+p_2)V^{u_3 \beta \rho}_{a_3 d e}(p_3 ,q+p_2 ,-q-p_3 ) \nonumber \\
& & G_{\rho \sigma}(q+p_2 +p_3 )V^{u_1 u_4 \sigma \mu}_{a_1 a_4 f a} 
\bigg{\}}\delta^{2\omega}(p_1 +p_2 +p_3 +p_4 ) .\nonumber 
\end{eqnarray}
\begin{figure}[h]
\begin{center}
\input{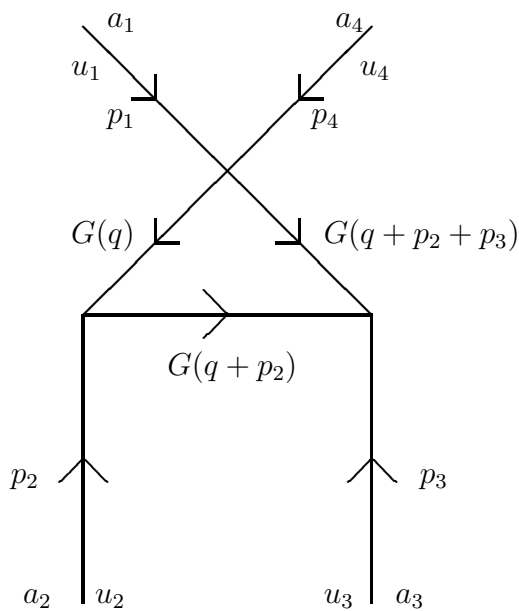}	
\end{center}
\caption{ \em Complete Lynx Diagram}
\label{fig:lynx0}
\end{figure}

The propagators $G_{\gamma \delta}$ are again split according to
Eq. (\ref{eq:lcprop}), giving eight lynx ``subdiagrams'', the simplest being
the diagram with {\em no} noncovariant factors (Fig.~\ref{fig:lynx0lc}), namely
\begin{eqnarray}
\label{eq:3p0lc}
J^{lynx}_{0}= \bigg{\{ } & & \int \frac{d^{2 \omega}q}{(2\pi)^{2\omega}}{\delta}^{ab}{\delta}^{cd}{\delta}^{ef}\\
& & G^{NLC}_{\mu \nu}(q)V^{u_2 \nu \alpha}_{a_2 b c}(p_2 ,q, -q-p_2 ) \nonumber \\
& & G^{NLC}_{\alpha \beta}(q+p_2)V^{u_3 \beta \rho}_{a_3 d e}(p_3 ,q+p_2 ,-q-p_3 ) \nonumber \\
& & G^{NLC}_{\rho \sigma}(q+p_2 +p_3 )V^{u_1 u_4 \sigma \mu}_{a_1 a_4 f a} 
\bigg{\} }\delta^{2\omega}(p_1 +p_2 +p_3 +p_4 ) ,\nonumber 
\end{eqnarray}
\begin{figure}[h]
\begin{center}
\input{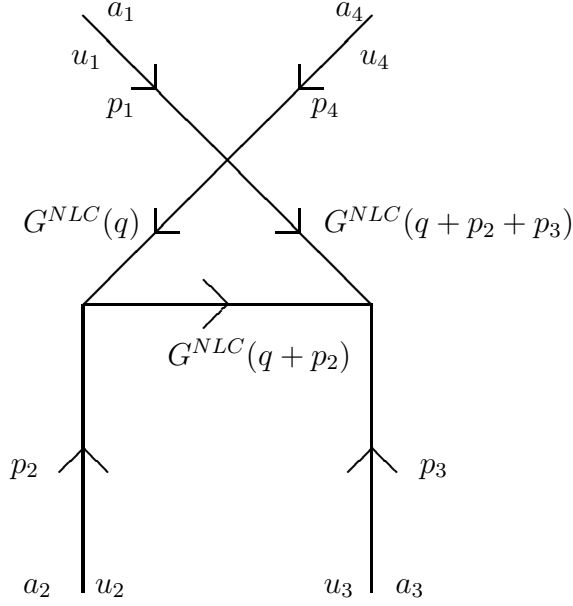}
\end{center}
\caption{ \em Lynx Diagram: No Noncovariant Propagators}
\label{fig:lynx0lc}
\end{figure}
which leads to the simple expression, 
\begin{eqnarray}
J^{lynx}_{0} = -\frac {\overline{I}}{4}\,\bigg{\{}
f^{a_1 ab}f^{a_2 bc} f^{a_3 cd}f^{a_4 da}(
2\,g_{u_1 u_2}g_{u_3 u_4} 
&+&2\,g_{u_1 u_3}g_{u_2 u_4} \\
&+& 26\,g_{u_1 u_4}g_{u_2 u_3})\nonumber \\
+\frac{N}{2} f^{a_1 a_4 e}f^{a_2 a_3 e}(
8\,g_{u_1 u_2}g_{u_3 u_4} 
&-& 10\,g_{u_1 u_3}g_{u_2 u_4} \nonumber \\
&-& 13\,g_{u_1 u_4}g_{u_2 u_3}) \bigg{\}}.\nonumber 
\end{eqnarray}

Using the same approach as in the reduction of the complete box diagram
(Sec.~\ref{sec:bxsec}), we now proceed to compute the remaining sets of
lynx ``subdiagrams'', containing, respectively a single $G^{LC}$ propagator,
and two and three $G^{LC}$ propagators. For {\em one} $G^{LC}$ propagator,
we get the three lynx ``subdiagrams'' depicted in Fig.~\ref{fig:lynx1lc}. The
corresponding integral $J^{lynx}_1$ is given by
\begin{eqnarray}
\label{eq:3p1lc}
J^{lynx}_{1}=\bigg{\{ } & & \int \frac{d^{2 \omega}q}{(2\pi)^{2\omega}}{\delta}^{ab}{\delta}^{cd}{\delta}^{ef}
G^{LC}_{\mu \nu}(q)V^{u_2 \nu \alpha}_{a_2 b c}(p_2 ,q, -q-p_2 ) \nonumber \\
& & G^{NLC}_{\alpha \beta}(q+p_2)V^{u_3 \beta \rho}_{a_3 d e}(p_3 ,q+p_2 ,-q-p_3 ) \nonumber \\
& & G^{NLC}_{\rho \sigma}(q+p_2 +p_3 )V^{u_1 u_4 \sigma \mu}_{a_1 a_4 f a} \\
&+& (1\rightarrow 4, 2\rightarrow 3)\bigg{\}} 
\delta^{2\omega}(p_1 +p_2 +p_3 +p_4 ) \nonumber \\
+ \bigg{\{}& & \int \frac{d^{2\omega}q}{(2\pi)^{2\omega}}{\delta}^{ab}{\delta}^{cd}{\delta}^{ef}
G^{NLC}_{\mu \nu}(q)V^{u_2 \nu \alpha}_{a_2 b c}(p_2 ,q, -q-p_2 ) \nonumber \\
& & G^{LC}_{\alpha \beta}(q+p_2)V^{u_3 \beta \rho}_{a_3 d e}(p_3 ,q+p_2 ,-q-p_3 ) \nonumber \\
& & G^{NLC}_{\rho \sigma}(q+p_2 +p_3 )V^{u_1 u_4 \sigma \mu}_{a_1 a_4 f a}\bigg{\}}
\delta^{2\omega}(p_1 +p_2 +p_3 +p_4 ),\nonumber 
\end{eqnarray}
\begin{figure}[h]
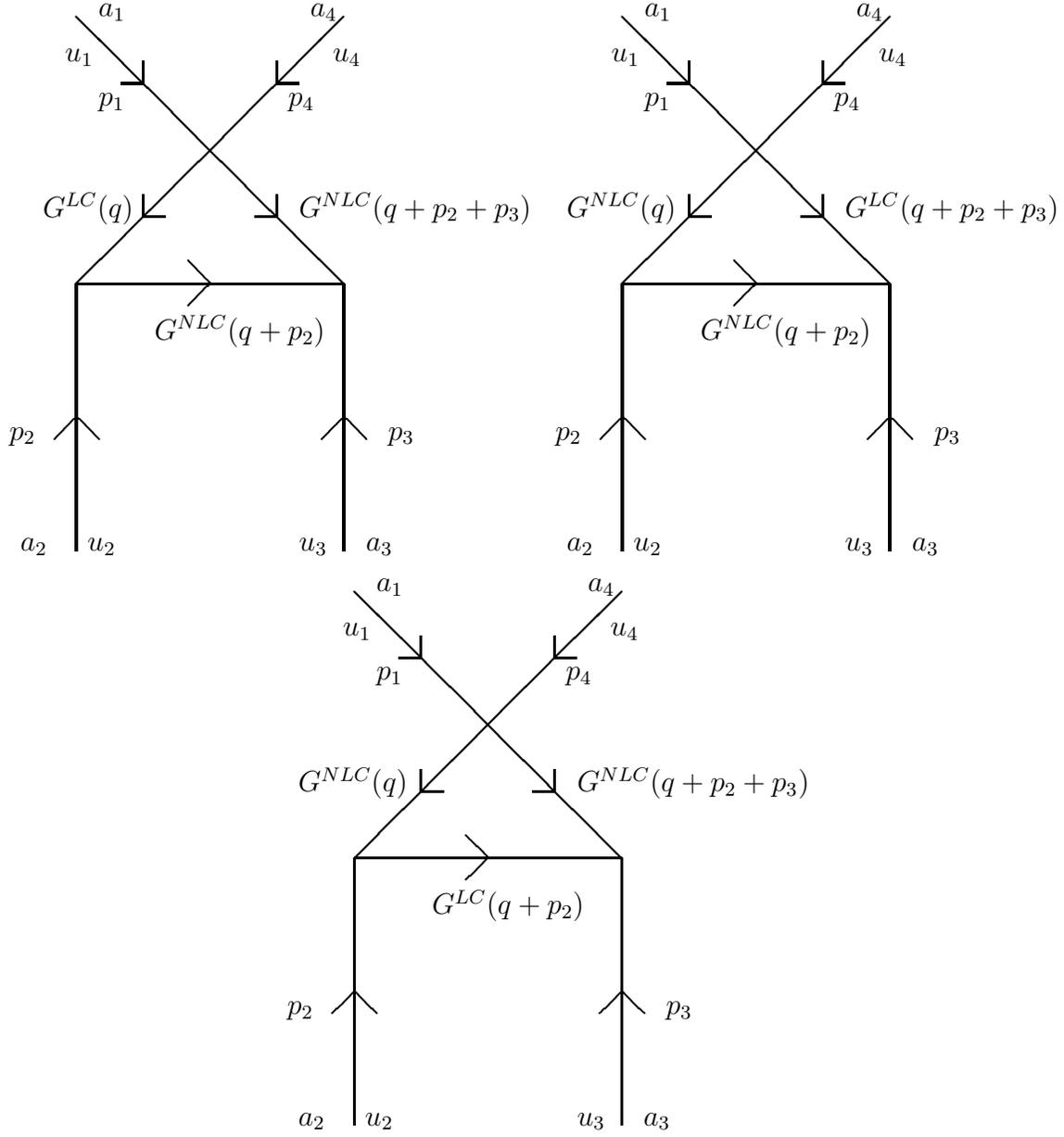

\begin{center}
\input{lynx1lc-a.te}
\vskip .1in
\input{lynx1lc-b.te}
\end{center}
\caption{ \em Lynx Diagrams: One Noncovariant Propagator}
\label{fig:lynx1lc}
\end{figure}
whose pole part is
\begin{eqnarray}
\hspace{-.2in} J^{lynx}_{1} \hspace{-.1in} = \frac {\overline{I}}{4\,(n\tdot n^{\ast})^{2}}\,
\bigg{\{}& &\hspace{-.2in} f^{a_1 ab}f^{a_2 bc} f^{a_3 cd}f^{a_4 da}\big[
3\,(n\tdot n^{\ast})^{2}(g_{u_1 u_2}g_{u_3 u_4}
+\,g_{u_1 u_3}g_{u_2 u_4}) \\
&+&(n\tdot n^{\ast})(3\,n_{u_1}\,n^{\ast}_{u_3}g_{u_2 u_4}
+3\,n_{u_4}\,n^{\ast}_{u_3}g_{u_1 u_2}
-9\,n_{u_1}\,n^{\ast}_{u_4}g_{u_2 u_3}\nonumber \\
&-&9\,n_{u_4}\,n^{\ast}_{u_1}g_{u_2 u_3}
-5\,n_{u_1}\,n^{\ast}_{u_2}g_{u_3 u_4}
- 5\,n_{u_4}\,n^{\ast}_{u_2}g_{u_1 u_3}\nonumber \\
&+&4\,n_{u_3}\,n^{\ast}_{u_1}g_{u_2 u_4}
+4\,n_{u_3}\,n^{\ast}_{u_4}g_{u_1 u_2}
-2\,n_{u_2}\,n^{\ast}_{u_1}g_{u_3 u_4}\nonumber \\
&-&2\,n_{u_2}\,n^{\ast}_{u_4}g_{u_1 u_3}
-12\,n_{u_3}\,n^{\ast}_{u_2}g_{u_1 u_4}
-14\,n_{u_2}\,n^{\ast}_{u_3}g_{u_1 u_4}) \nonumber \\
&+&n_{u_3}\,n^{\ast}_{u_2}n^{\ast}_{u_4}n_{u_1}
+n_{u_4}\,n^{\ast}_{u_1}n^{\ast}_{u_2}n_{u_3}
+3\,n_{u_4}\,n^{\ast}_{u_1}n^{\ast}_{u_3}n_{u_2}\nonumber \\
&+&3\,n_{u_2}\,n^{\ast}_{u_3}n^{\ast}_{u_4}n_{u_1}
+10\,n_{u_4}\,n^{\ast}_{u_2}n^{\ast}_{u_3}n_{u_1}
-6\,n_{u_3}\,n^{\ast}_{u_1}n^{\ast}_{u_4}n_{u_2}\big]
\nonumber \\
+\frac{N}{2} f^{a_1 a_4 e}f^{a_2 a_3 e}\big[ & & \hspace{-.1in}
(n\tdot n^{\ast})^{2}(18\,g_{u_1 u_4}g_{u_2 u_3}
-9\,g_{u_1 u_4}g_{u_2 u_3}
-3\,g_{u_1 u_3}g_{u_2 u_4})\nonumber \\
&+&(n\tdot n^{\ast})(3\,n_{u_4}\,n^{\ast}_{u_1}g_{u_2 u_3}
+6\,n_{u_1}\,n^{\ast}_{u_4}g_{u_2 u_3} 
+6\,n_{u_3}\,n^{\ast}_{u_2}g_{u_1 u_4}\nonumber \\
&-&2\,n_{u_3}\,n^{\ast}_{u_1}g_{u_2 u_4}
-2\,n_{u_3}\,n^{\ast}_{u_4}g_{u_1 u_2}
+7\,n_{u_2}\,n^{\ast}_{u_1}g_{u_3 u_4}\nonumber \\
&+&7\,n_{u_2}\,n^{\ast}_{u_3}g_{u_1 u_4}
+7\,n_{u_4}\,n^{\ast}_{u_2}g_{u_1 u_3}
-5\,n_{u_2}\,n^{\ast}_{u_4}g_{u_1 u_3}\nonumber \\
&-&2\,n_{u_1}\,n^{\ast}_{u_2}g_{u_3 u_4}
-18\,n_{u_1}\,n^{\ast}_{u_3}g_{u_2 u_4}
+15\,n_{u_4}\,n^{\ast}_{u_3}g_{u_1 u_2})\nonumber \\
&+&3\,n_{u_3}\,n^{\ast}_{u_1}n^{\ast}_{u_4}n_{u_2}
+3\,n_{u_2}\,n^{\ast}_{u_3}n^{\ast}_{u_4}n_{u_1}
+4\,n_{u_3}\,n^{\ast}_{u_2}n^{\ast}_{u_4}n_{u_1}\nonumber \\
&-&5\,n_{u_4}\,n^{\ast}_{u_1}n^{\ast}_{u_2}n_{u_3}
-5\,n_{u_4}\,n^{\ast}_{u_2}n^{\ast}_{u_3}n_{u_1}
-6\,n_{u_4}\,n^{\ast}_{u_1}n^{\ast}_{u_3}n_{u_2}
\big]\bigg{\}}. \nonumber
\end{eqnarray}

The next ``subdiagrams'' are those containing {\em two} propagators 
$G^{LC}$ (Fig.~\ref{fig:lynx2lc}), the associated Feynman integral 
being $J^{lynx}_2$, where
\begin{figure}[h]
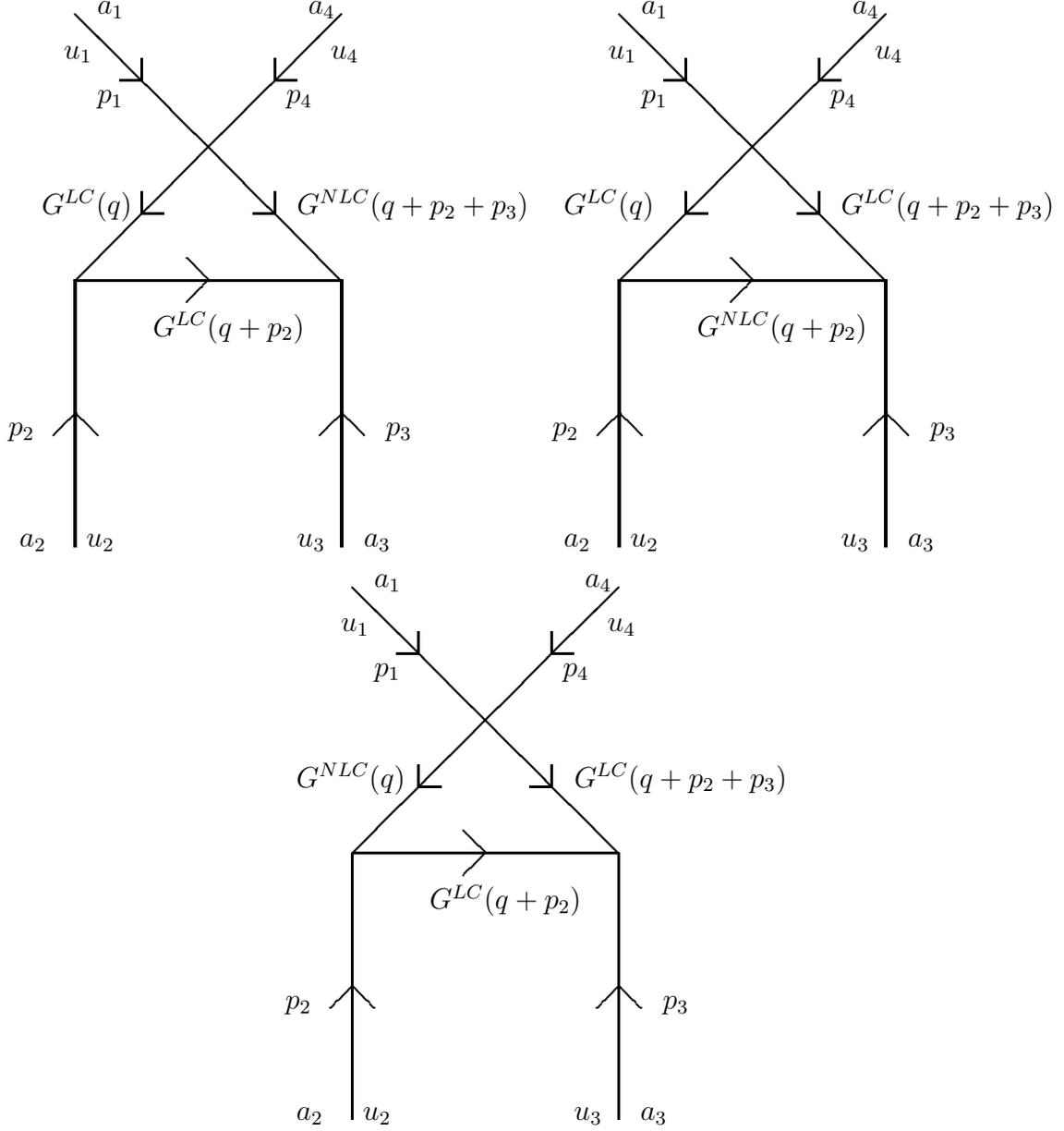

\begin{center}
\input{lynx2lc-a.te}
\vskip .1in
\input{lynx2lc-b.te}
\end{center}
\caption{ \em Lynx Diagrams: Two Noncovariant Propagators}
\label{fig:lynx2lc}
\end{figure}
\begin{eqnarray}
\label{eq:3p2lc}
J^{lynx}_{2}= \bigg{\{ } & & \int \frac{d^{2 \omega}q}{(2\pi)^{2\omega}}{\delta}^{ab}{\delta}^{cd}{\delta}^{ef}
G^{LC}_{\mu \nu}(q)V^{u_2 \nu \alpha}_{a_2 b c}(p_2 ,q, -q-p_2 ) \\
& & G^{LC}_{\alpha \beta}(q+p_2)V^{u_3 \beta \rho}_{a_3 d e}(p_3 ,q+p_2 ,-q-p_3 ) \nonumber \\
& & G^{NLC}_{\rho \sigma}(q+p_2 +p_3 )V^{u_1 u_4 \sigma \mu}_{a_1 a_4 f a} \nonumber \\
&+& (1\rightarrow 4, 2\rightarrow 3)\bigg{\}}\delta^{2\omega}(p_1 +p_2 +p_3 +p_4 ) \nonumber \\
+\bigg{\{}& & \int \frac{d^{2 \omega}q}{(2\pi)^{2\omega}}{\delta}^{ab}{\delta}^{cd}{\delta}^{ef}
G^{LC}_{\mu \nu}(q)V^{u_2 \nu \alpha}_{a_2 b c}(p_2 ,q, -q-p_2 ) \nonumber \\
& & G^{NLC}_{\alpha \beta}(q+p_2)V^{u_3 \beta \rho}_{a_3 d e}(p_3 ,q+p_2 ,-q-p_3 ) \nonumber \\
& & G^{LC}_{\rho \sigma}(q+p_2 +p_3 )V^{u_1 u_4 \sigma \mu}_{a_1 a_4 f a}
\bigg{\} }\delta^{2\omega}(p_1 +p_2 +p_3 +p_4 ).\nonumber 
\end{eqnarray}
The integrated UV divergent contribution to the expression is two pages long and is stated in Appendix A,
Eq.~(\ref{eq:lx2lc}).

There is a final, single lynx ``subdiagram'' with {\em three} noncovariant
propagators (Fig.~\ref{fig:lynx3lc}), which leads to the formula
\begin{eqnarray}
\label{eq:3p3lc}
J^{lynx}_{3}= \bigg{\{ } & & \int \frac{d^{2 \omega}q}{(2\pi)^{2\omega}}{\delta}^{ab}{\delta}^{cd}{\delta}^{ef}
G^{LC}_{\mu \nu}(q)V^{u_2 \nu \alpha}_{a_2 b c}(p_2 ,q, -q-p_2 ) \\
& & G^{LC}_{\alpha \beta}(q+p_2)V^{u_3 \beta \rho}_{a_3 d e}(p_3 ,q+p_2 ,-q-p_3 ) \nonumber \\
& & G^{LC}_{\rho \sigma}(q+p_2 +p_3 )V^{u_1 u_4 \sigma \mu}_{a_1 a_4 f a} 
\bigg{\} }\delta^{2\omega}(p_1 +p_2 +p_3 +p_4 ).\nonumber 
\end{eqnarray}
\begin{figure}[h]
\begin{center}
\input{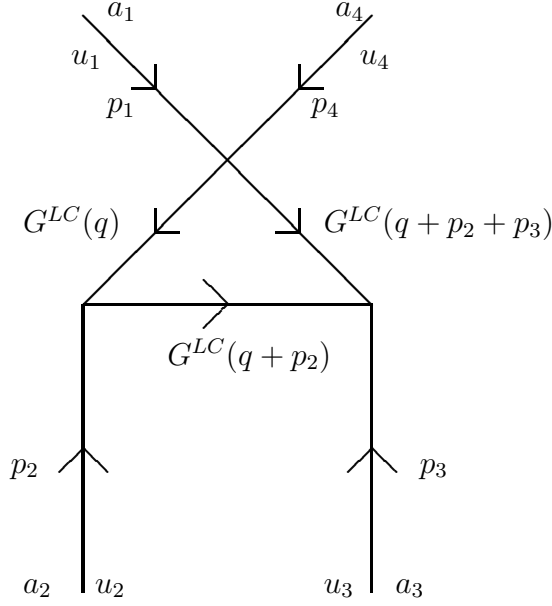}
\end{center}
\caption{ \em Lynx Diagram: Three Noncovariant Propagators}
\label{fig:lynx3lc}
\end{figure}
We refrain from listing the integrated answer.

\subsection{Fish Diagrams} \label{sec:fshsec}
The last, yet easiest, diagram to compute is the complete fish diagram,
shown in Fig.~\ref{fig:fish0}. It consists of only two four-point vertices
and is given by
\begin{eqnarray} 
\label{eq:fisheqn}
J^{fish}= \bigg{\{ } & & \int \frac{d^{2 \omega}q}{(2\pi)^{2\omega}}{\delta}^{ab}{\delta}^{cd}
G_{\mu \nu}(q)V^{u_1 u_4 \beta \mu}_{a_1 a_4 d a}\\
& & G_{\alpha \beta}(q+p_2 +p_3)V^{u_3 u_2 \nu \alpha}_{a_3 a_2 \nu \alpha}\bigg{\}}
\delta^{2\omega}(p_1 +p_2 +p_3 +p_4 ).\nonumber 
\end{eqnarray}
\begin{figure}[h]
\begin{center}
\input{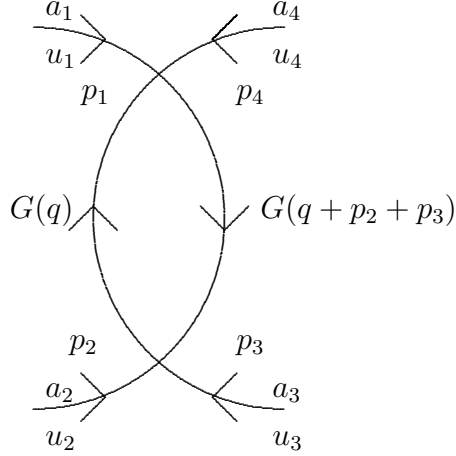}
\end{center}
\caption{ \em Complete Fish Diagram}
\label{fig:fish0}
\end{figure}
Separation of the gauge propagator $G^{ab}_{\mu\nu}$ into covariant
and noncovariant propagators (as in Eq.(~\ref{eq:lcprop})) produces four
``subdiagrams''. The Feynman integral for the strictly {\em covariant}
fish diagram (Fig.~\ref{fig:fish0lc}) reads
\begin{eqnarray}
\label{eq:2p0lc}
J^{fish}_{0}= \delta^{2\omega}(p_1 +p_2 +p_3 +p_4 ) \bigg{\{ } & & 
\int \frac{d^{2 \omega}q}{(2\pi)^{2\omega}}{\delta}^{ab}{\delta}^{cd}
G^{NLC}_{\mu \nu}(q)V^{u_1 u_4 \beta \mu}_{a_1 a_4 d a}\\
& & 
G^{NLC}_{\alpha \beta}(q+p_2 +p_3 )V^{u_3 u_2 \nu \alpha}_{a_3 a_2 b c}
\bigg{\} } \nonumber \\
=\overline{I}\; \bigg{\{} f^{a_1 ab}f^{a_2 bc} f^{a_3 cd}f^{a_4 da}
 \big[& & 
g_{u_1 u_2}\, g_{u_3 u_4}\ +\ g_{u_1 u_3}\, g_{u_2 u_4} \nonumber\\
&+& 
4\,g_{u_1 u_4}\, g_{u_2 u_3}\big] \nonumber \\ 
+ \frac{N}{2} f^{a_1 a_4 e}f^{a_2 a_3 e} \big[ & & 4\,g_{u_1 u_2}\, g(u_3 ,u_4 ) 
- 5\,g_{u_1 u_3}\, g_{u_2 u_4}\nonumber \\
&-& 
2\,g_{u_1 u_4} g_{u_2 u_3}\big]\bigg{\}}. \nonumber 
\end{eqnarray}
\begin{figure}[h]
\begin{center}
\input{fish0lc.pictex.te}
\end{center}
\vskip .1in
\caption{ \em Fish Diagram: No Noncovariant Propagators}
\label{fig:fish0lc}
\vskip .1in
\end{figure}

There are two fish ``subdiagrams'' (Fig.~\ref{fig:fish1lc}) having {\em
one} noncovariant factor each, namely
\begin{eqnarray}
\label{eq:2p1lc}
J^{fish}_{1}=\delta^{2\omega}(p_1 +p_2 +p_3 +p_4 ) \bigg{\{ } & & \int \frac{d^{2 \omega}q}{(2\pi)^{2\omega}}{\delta}^{ab}{\delta}^{cd}
G^{LC}_{\mu \nu}(q)V^{u_4 u_1 \beta \mu}_{a_1 a_4 c a} \\
& & G^{NLC}_{\alpha \beta}(q+p_2 +p_3 )
V^{u_3 u_2 \nu \alpha}_{a_3 a_2 b c} \nonumber \\
&+& (1 \leftrightarrow 4, 2\leftrightarrow 3)
\bigg{\}} .\nonumber 
\end{eqnarray}
\begin{figure}[h]
\begin{center}
\input{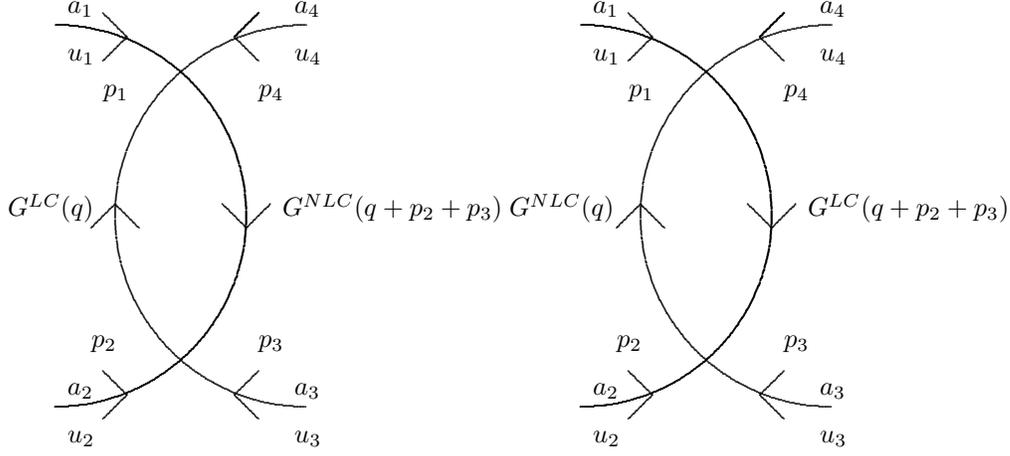}
\end{center}
\caption{ \em Fish Diagrams: One Noncovariant Propagator}
\label{fig:fish1lc}
\vskip .1in
\end{figure}
Integration of Eq.~(\ref{eq:2p1lc}) yields the following UV contribution
\begin{eqnarray}
J^{fish}_{1}=
-\frac{\overline{I}}{2(n\tdot n^{\ast})}
\,\bigg{\{} \big[f^{a_1 ab}f^{a_2 bc} f^{a_3 cd}f^{a_4 da}
g_{u_1 u_2}\,n_{u_3 }n^{\ast}_{u_4 }
&+&g_{u_1 u_2}\,n_{u_4 }n^{\ast}_{u_3 }\mbox{} \\
+g_{u_1 u_3}\,n_{u_2 }n^{\ast}_{u_4 } 
&+&g_{u_1 u_3}\,n_{u_4 }n^{\ast}_{u_2 }\mbox{} \nonumber \\
+g_{u_2 u_4}\,n_{u_3 }n^{\ast}_{u_1 } 
&+&g_{u_2 u_4}\,n_{u_1 }n^{\ast}_{u_3 }\mbox{} \nonumber \\
+g_{u_3 u_4}\,n_{u_1 }n^{\ast}_{u_2 }
&+&g_{u_3 u_4}\,n_{u_2 }n^{\ast}_{u_1 }\mbox{} \nonumber \\
-4\,g_{u_2 u_3}\,n_{u_1 }n^{\ast}_{u_4 }
&+&\,g_{u_2 u_3}\,n_{u_4 }n^{\ast}_{u_1 }\mbox{} \nonumber \\
+\,g_{u_1 u_4}\,n_{u_3 }n^{\ast}_{u_2 } 
&+&\,g_{u_1 u_4}\,n_{u_2 }n^{\ast}_{u_3 })\mbox{} \nonumber \\
+8\,(n \tdot n^{\ast})& & \hspace{-.15in} g_{u_1 u_4}g_{u_2 u_3}\big]
\mbox{} \nonumber \\
+\frac{N}{2} f^{a_1 a_4 e}f^{a_2 a_3 e}\big[
2\,(g_{u_2 u_3}\,n_{u_1 }n^{\ast}_{u_4 }
&+&\,g_{u_2 u_3}\,n_{u_4 }n^{\ast}_{u_1 }\mbox{} \nonumber \\
+\,g_{u_1 u_4}\,n_{u_3 }n^{\ast}_{u_2 } 
&+&\,g_{u_1 u_4}\,n_{u_2 }n^{\ast}_{u_3 }) \mbox{} \nonumber \\
+4\,(g_{u_1 u_2}\,n_{u_3 }n^{\ast}_{u_4 }
&+&\,g_{u_1 u_2}\,n_{u_4 }n^{\ast}_{u_3 }\mbox{} \nonumber \\
+\,g_{u_3 u_4}\,n_{u_1 }n^{\ast}_{u_2 } 
&+&\,g_{u_3 u_4}\,n_{u_2 }n^{\ast}_{u_1 })\mbox{} \nonumber \\
-5\,(g_{u_1 u_3}\,n_{u_2 }n^{\ast}_{u_4 }
&+&\,g_{u_1 u_3}\,n_{u_4 }n^{\ast}_{u_2 }\mbox{} \nonumber \\
+\,g_{u_2 u_4}\,n_{u_3 }n^{\ast}_{u_1 } 
&+&\,g_{u_2 u_4}\,n_{u_1 }n^{\ast}_{u_3 })\mbox{} \nonumber \\ 
-4\,(n \tdot n^{\ast})& & \hspace{-.15in} g_{u_1 u_4}g_{u_2 u_3}\big]
\bigg{\}}\mbox{} \nonumber \\
-\frac {\overline{I}}{2\,(n\tdot n^{\ast})}
\bigg{\{} f^{a_1 ab}f^{a_2 bc} f^{a_3 cd}f^{a_4 da}(
g_{u_1 u_2}\,n_{u_3 }n^{\ast}_{u_4 }
&+&g_{u_1 u_3}\,n_{u_2 }n^{\ast}_{u_4 }\mbox{} \nonumber \\
+g_{u_1 u_3}\,n_{u_4 }n^{\ast}_{u_2 }
&+&g_{u_1 u_2}\,n_{u_4 }n^{\ast}_{u_3 }\mbox{} \nonumber \\
+g_{u_2 u_4}\,n_{u_3 }n^{\ast}_{u_1 }
&+&g_{u_2 u_4}\,n_{u_1 }n^{\ast}_{u_3 }\mbox{} \nonumber \\
+g_{u_3 u_4}\,n_{u_1 }n^{\ast}_{u_2 }
&+&g_{u_3 u_4}\,n_{u_2 }n^{\ast}_{u_1 }\mbox{} \nonumber \\
-4\,(g_{u_2 u_3}\,n_{u_1 }n^{\ast}_{u_4 }
&+&\,g_{u_2 u_3}\,n_{u_4 }n^{\ast}_{u_1 }\mbox{} \nonumber \\
+\,g_{u_1 u_4}\,n_{u_3 }n^{\ast}_{u_2 }
&+&4\,g_{u_1 u_4}\,n_{u_2 }n^{\ast}_{u_3 })\mbox{} \nonumber \\
+8\,(n\tdot n^{\ast})& & \hspace{-.15in} g_{u_1 u_4}g_{u_2 u_3})
\mbox{} \nonumber \\
+\frac{N}{2} f^{a_1 a_4 e}f^{a_2 a_3 e}\big[
2\,(g_{u_2 u_3}\,n_{u_1 }n^{\ast}_{u_4 }
&+&\,g_{u_2 u_3}\,n_{u_4 }n^{\ast}_{u_1 }\mbox{} \nonumber \\
+\,g_{u_1 u_4}\,n_{u_3 }n^{\ast}_{u_2 }
&+&\,g_{u_1 u_4}\,n_{u_2 }n^{\ast}_{u_3 })\mbox{} \nonumber \\
+4\,(g_{u_1 u_2}\,n_{u_3 }n^{\ast}_{u_4 }
&+&\,g_{u_1 u_2}\,n_{u_4 }n^{\ast}_{u_3 }\mbox{} \nonumber \\
+\,g_{u_3 u_4}\,n_{u_1 }n^{\ast}_{u_2 }
&+&\,g_{u_3 u_4}\,n_{u_2 }n^{\ast}_{u_1 })\mbox{} \nonumber \\
-5\,(g_{u_1 u_3}\,n_{u_2 }n^{\ast}_{u_4 }
&+&\,g_{u_1 u_3}\,n_{u_4 }n^{\ast}_{u_2 }\mbox{} \nonumber \\
+\,g_{u_2 u_4}\,n_{u_3 }n^{\ast}_{u_1 }
&+&\,g_{u_2 u_4}\,n_{u_1 }n^{\ast}_{u_3 })\mbox{} \nonumber \\
-4\,(n\tdot n^{\ast})& & \hspace{-.15in} g_{u_1 u_4}g_{u_2 u_3}) \big]
\bigg{\}}\mbox{}.\nonumber
\end{eqnarray}

There remains a single fish ``subdiagram'' with {\em two} noncovariant
propagators (cf. Fig.~\ref{fig:fish2lc}), giving the contribution
\begin{eqnarray}
\label{eq:2p2lc}
J^{fish}_{2}=\delta^{2\omega}(p_1 +p_2 +p_3 +p_4 ) \bigg{\{ } & & \int \frac{d^{2 \omega}q}{(2\pi)^{2\omega}}{\delta}^{ab}{\delta}^{cd}
G^{LC}_{\mu \nu}(q)V^{u_1 u_4 \beta \mu}_{a_1 a_4 c a} \\
& & G^{LC}_{\alpha \beta}(q+p_2 +p_3 )V^{u_3 u_2 \nu \alpha}_{a_3 a_2 b c}
\bigg{\} } .\nonumber 
\end{eqnarray}
\begin{figure}[h]
\begin{center}
\input{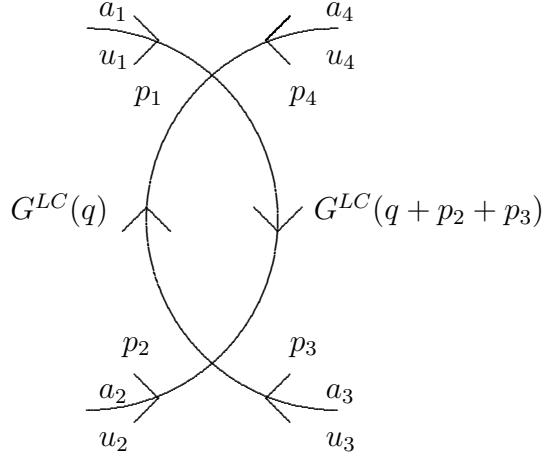}
\end{center}
\caption{ \em Fish Diagram: Two Noncovariant Propagators}
\label{fig:fish2lc}
\end{figure}
The answer is again lengthy and has been banished to Appendix A,
Eq.~(\ref{eq:fsh2lc}).

\clearpage
\clearpage
\section{Discussion}
\label{sec:disc}
Before presenting the complete result for the UV divergent contribution to
the 1PI four-point function in the light-cone gauge, we shall discuss
the constraints on the form of this result.

The presence of nonlocal terms in loop integrals complicates the
renormalization procedure appreciably. One could imagine a host of
nonlocal expressions that evade the Ward identities. In fact, at first
glance, it would seem that there could be an infinite number of such
expressions. Fortunately, there are additional constraints on the
structure of the non-polynomial divergences in the light-cone gauge
which conspire to exclude all but a few of such terms. As a
result, only a finite number of nonlocal counterterms is required to
renormalize the 1PI fuctions. Actually, these nonlocal couterterms
do not contribute to the complete Green functions and hence they do not
affect the observables of the theory \cite{Bas:4pt}.

We recall that any light-cone gauge expression can be assigned two
degrees of divergence, $\alpha$ and $ {\alpha}_{\perp}$, where $\alpha$
is the degree with respect to the total large-momentum behaviour of a
Feynman integral, while $\alpha_{\perp}$ is the degree with respect to
the transverse large-momentum behaviour. If both $\alpha$ and
$\alpha_{\perp}$ are negative for any diagram and its possible
``subdiagrams'', then the corresponding integral will be convergent. To see
this, consider the gauge propagator in Eq.~(\ref{eq:prop}): since
application of the derivative operator $\partial_{\perp}$ lowers both
$\alpha$ and $\alpha_{\perp}$, {\em repeated} use of $\partial_{\perp}$
on a {\em divergent} integral will eventually yield a {\em convergent}
one. We conclude, therefore, that the pole part of a subtracted 1PI
diagram (i.e. one in which the subdivergences have already been
subtracted) will be polynomial in $p_{\perp i}$, the transverse external
momenta.

It is furthermore known that in space-like axial gauges ($n^2<0$) only
{\em local} divergences are encountered in subtracted 1PI graphs, as
demonstrated in Appendix A of \cite{K1}. The reasoning is based on the
ability to write $q^{\mu}/n \tdot q$ as $\partial / \partial n_{\mu}\;
ln(q\tdot n)$, Wick rotate the resulting integral and apply Weinberg's
theorem \cite{Wein:th}. However, this reasoning does not hold in the
case of the light-cone gauge. A logarithmic representation of the
spurious denominator $(q \tdot n)^{-1}$ implies the principal-value
prescription, which is incompatible with a Wick rotation, and hence with
the $n^{\ast}_{\mu}$-prescription \cite{ns:L3}.

Although the $n^{\ast}_{\mu}$-prescription had originally been used to
handle the poles of $(q\tdot n)^{-1}$ in the light-cone gauge, the
$n^{\ast}_{\mu}$-prescription can be used with any values of $n_0$ and
${\bf n}$, provided that $|{\bf n}|\neq 0$. Thus, the space-like limit
$n_0 \rightarrow 0$, with ${\bf n}$ constant, which yields the space-like
axial gauge, is allowed when the $n^{\ast}_{\mu}$-prescription is used.
We conclude that the {\em nonlocal} divergent parts of a subtracted 1PI
diagram will become {\em local}, and perhaps even vanish for $n_0
\rightarrow 0$.

The constraints above are the basis for the prediction of the nonlocal
part of 1PI the four-point function by Bassetto, Dalbosco and Soldati
\cite{Bas:4pt}. Our explicit calculation yields the same tensor structure
as predicted by the authors of Ref \cite{Bas:4pt}, along with the local,
covariant part of the four-point function: 
\begin{eqnarray}
\label{eq:result}
{\Gamma}^{a_1 a_2 a_3 a_4}_{u_1 u_2 u_3 u_4}(p_1, p_2, p_3, p_4 )= 
& & -g^{2}\overline{I} \frac{2 N}{n\tdot n^{\ast}} f^{a_1 a_4 e}f^{a_2 a_3 e}\delta^{2\omega}(p_1 +p_2 +p_3 +p_4 ) \\
& & \hspace{-.2in} \bigg{\{}\;
\overline{n}^{\ast}_{\{u_1 }n^{}_{u_3 \}}g_{u_2 u_4 }
+\overline{n}^{\ast}_{\{u_2 }n^{}_{u_4 \}}g_{u_1 u_3}\nonumber \\
& & \hspace{-.2in} - \;\overline{n}^{\ast}_{\{u_2 }n^{}_{u_3 \}}g_{u_1 u_4}
-\overline{n}^{\ast}_{\{u_1 }n^{}_{u_4 \}}g_{u_2 u_3}\nonumber \\
& & \hspace{-.2in} +\;n_{u_1}n_{u_2}g_{u_3 u_4}
\frac{n\tdot (p_4 -p_3 )}{n\tdot (p_4 +p_3 )}
\bigg[\frac{n^{\ast}\tdot p_1 }{n\tdot p_1 }-\frac{n^{\ast}\tdot p_2 }{n\tdot p_2 }\bigg]
\nonumber \\
& & \hspace{-.2in} +\;n_{u_3 }n_{u_4 }g_{u_1 u_2}
\frac{n\tdot (p_2 -p_1 )}{n\tdot (p_2 +p_1 )}
\bigg[\frac{n^{\ast}\tdot p_3 }{n\tdot p_3 }-\frac{n^{\ast}\tdot p_4 }{n\tdot p_4 }\bigg]
\nonumber \\
& &  \hspace{-.2in} +\;2\big( n_{u_1 }{p_1}_{u_2 }-n_{u_2}{p_2}_{u_1 }\big)
\frac{n_{u_3}n_{u_4}}{n\tdot (p_3 +p_4 )}
\bigg[\frac{n^{\ast}\tdot p_4 }{n\tdot p_4 }-\frac{n^{\ast}\tdot p_3 }{n\tdot p_3 }\bigg]
\nonumber \\
& &  \hspace{-.2in} +\;2\big( n_{u_3}{p_3}_{u_4}-n_{u_4 }{p_4}_{u_3 }\big)
\frac{n_{u_1 }n_{u_2 }}{n\tdot (p_1 +p_2 )}
\bigg[\frac{n^{\ast}\tdot p_2 }{n\tdot p_2 }-\frac{n^{\ast}\tdot p_1 }{n\tdot p_1 }\bigg]
\nonumber \\
& &  \hspace{-.2in} +\;n_{u_1 }n_{u_2 }n_{u_3 }n_{u_4 }
\bigg[\bigg(
\frac{p_2 ^{2}}{n\tdot p_2 }-\frac{p_1 ^{2}}{n\tdot p_1 }\bigg)
\frac{1}{n\tdot (p_3 +p_4 )}
\bigg(\frac{n^{\ast}\tdot p_4 }{n\tdot p_4 }-\frac{n^{\ast}\tdot p_3 }{n\tdot p_3 }\bigg)
\nonumber \\
& & \hspace{-.2in} +\;\bigg(\frac{p_4 ^{2}}{n\tdot p_4 }-\frac{p_3 ^{2}}{n\tdot p_3 }\bigg)
\frac{1}{n\tdot (p_1 +p_2 )}
\bigg(\frac{n^{\ast}\tdot p_2 }{n\tdot p_2 }-\frac{n^{\ast}\tdot p_1 }{n\tdot p_1 }\bigg)\bigg]
\bigg{\}}\nonumber \\
& & \hspace{-.2in} +\;(2\rightarrow 3,3 \rightarrow 4,4 \rightarrow 2)\nonumber \\
& & \hspace{-.2in} +\;(2\leftrightarrow 3)\nonumber \\
& & \hspace{-.2in} +\;\frac{22 \overline{I}}{3}\bigg{\{}
g_{u_1 u_2}g_{u_3 u_4}\bigg( f^{a_1 a_4 e}f^{a_2 a_3 e}-f^{a_1 a_3 e}f^{a_4 a_2 e}\bigg)
\nonumber \\
& & \hspace{-.2in} +\;g_{u_1 u_3}g_{u_2 u_4}\bigg( f^{a_1 a_2 e}f^{a_3 a_4 e}-f^{a_1 a_4 e}f^{a_2 a_3 e}\bigg)
\nonumber \\
& & \hspace{-.2in} +\;g_{u_1 u_4}g_{u_2 u_3}\bigg( f^{a_1 a_3 e}f^{a_4 a_2 e}-f^{a_1 a_2 e}f^{a_3 a_4 e}\bigg)
\bigg{\}},\nonumber
\end{eqnarray}
where $\overline{n}^{\ast}_{u_i}=n^{\ast}_{u_i}-n_{u_i}n^{\ast}\tdot p_{i}/n\tdot p_{i}$.

As expected, all terms are polynomial in $p_{\perp i}$,
since the only non-polynomial factors are of the form
$(p\tdot n)^{-1}=(p_{\parallel}\tdot n)^{-1}$. Moreover, in the limit $n_0
\rightarrow 0$ all terms become {\em local}, since \begin{equation}
\frac{n^{\ast}\tdot p}{n\tdot p} {\bigg|}_{n_0 \rightarrow 0}=-1.
\end{equation}

Finally, the Ward identity
\begin{eqnarray}
ip^{u_1}_1 {\Gamma}^{a_1 a_2 a_3 a_4}_{u_1 u_2 u_3 u_4}(p_1 ,p_2 ,p_3 ,p_4 )
&=& 
-g f^{a_1 a_2 e}{\Gamma}^{a_3 a_4 e}_{u_2 u_3 u_4}(-p_3 -p_4 ,p_3 ,p_4 ) \\
\! & &\!\!
-g f^{a_1 a_3 e}{\Gamma}^{a_4 a_2 e}_{u_3 u_4 u_2}(-p_4 -p_2 ,p_4 ,p_2 )
\nonumber \\
\! & &\!\! 
-g f^{a_1 a_4 e}{\Gamma}^{a_2 a_3 e}_{u_4 u_2 u_3}(-p_2 -p_3 ,p_2 ,p_3 )
\nonumber 
\end{eqnarray}
has been explicitly verified including the covariant parts and
the relative weights, by using the 1PI three-point function presented in
\cite{Dal:3pt}.

It may be interesting to note that, although the colour factor 
$f^{a_1 ab}f^{a_2 bc}f^{a_3 cd}f^{a_4 da}$ and its permuted variations occur throughout the intermediate stages of calculation (in fact, these are the only colour factors that appear in the box diagrams), none of these terms survive in the final result (\ref{eq:result}).

\clearpage
\clearpage
     \begin{center}     {\bf {\large Appendix A: Intermediate Results}}     \end{center}

In this appendix we give some of the longer intermediate results of our box,
lynx and fish diagram calculations.

1). $J^{box}_{2}$, the integrated result of (\ref{eq:4p2lc}):
\begin{eqnarray}   \label{eq:bx2lc}
J^{box}_{2}=
%
%
\overline{I} f^{a_1 ab}f^{a_2 bc}f^{a_3 cd}f^{a_4 da} \bigg{\{}
\frac{1}{12}
\big[ 13\,g_{u_2 u_4}g_{u_1 u_3}
+ 4\,(g_{u_3 u_4} g_{u_1 u_2}\;\;\;\;\; & & \mbox{} \\
+g_{u_2 u_3} g_{u_1 u_4})\big]\;\;\;\;\; & & \mbox{} \nonumber \\
-\frac {5\,n_{u_1}}{(n\tdot n^{\ast})n\tdot (p_2 + p_3 +p_4 )}
\big[g_{u_2 u_3}n_{u_4}(n^{\ast}\tdot p_2 )
+g_{u_2 u_3}n_{u_4}(n^{\ast}\tdot p_3 ) \;\;\;\;\; & & \mbox{}\nonumber\\
+g_{u_2 u_3}n_{u_4}(n^{\ast}\tdot p_4 )
+n_{u_2}g_{u_3 u_4}(n^{\ast}\tdot p_2 )\;\;\;\;\; & & \mbox{}\nonumber \\
+n_{u_2}g_{u_3 u_4}(n^{\ast}\tdot p_3 )
+n_{u_2}g_{u_3 u_4}(n^{\ast}\tdot p_4 )\big]\;\;\;\;\; & & \mbox{}\nonumber \\
+\frac{1}{4n\tdot ( p_2 +p_3 +p_4 )(n\tdot n^{\ast})^{2}}
\big[ 3\,(n_{u_2}n^{\ast}_{u_1}n^{\ast}_{u_4}n_{u_3}(n\tdot p_2 )
+\,n_{u_4}n^{\ast}_{u_2}n^{\ast}_{u_1}n_{u_3}(n\tdot p_2 )\;\;\;\;\; & & \mbox{}\nonumber \\
+\,n_{u_2}n^{\ast}_{u_1}n^{\ast}_{u_4}n_{u_3}(n\tdot p_3 )
+\,n_{u_2}n^{\ast}_{u_1}n^{\ast}_{u_4}n_{u_3}(n\tdot p_4 )\;\;\;\;\; & & \mbox{}\nonumber \\
+\,n_{u_4}n^{\ast}_{u_2}n^{\ast}_{u_1}n_{u_3}(n\tdot p_3 )
+\,n_{u_4}n^{\ast}_{u_2}n^{\ast}_{u_1}n_{u_3}(n\tdot p_4 ))\;\;\;\;\; & & \mbox{}\nonumber \\
-2\,(n_{u_1}n^{\ast}_{u_2}n^{\ast}_{u_4}n_{u_3}(n\tdot p_2 
)+\,n_{u_1}n^{\ast}_{u_2}n^{\ast}_{u_4}n_{u_3}(n\tdot p_3 )\;\;\;\;\; & & \mbox{}\nonumber \\
+\,n_{u_1}n^{\ast}_{u_2}n^{\ast}_{u_4}n_{u_3}(n\tdot p_4 ))\;\;\;\;\; & & \mbox{}\nonumber \\
+6\,(n_{u_2}n^{\ast}_{u_1}n^{\ast}_{u_3}n_{u_4}(n\tdot p_2 )
+\,n_{u_2}n^{\ast}_{u_1}n^{\ast}_{u_3}n_{u_4}(n\tdot p_3 )\;\;\;\;\; & & \mbox{}\nonumber \\
+\,n_{u_2}n^{\ast}_{u_1}n^{\ast}_{u_3}n_{u_4}(n\tdot p_4 ))\;\;\;\;\; & & \mbox{} \nonumber \\
+13\,(n_{u_1}n_{u_4}n^{\ast}_{u_2}n^{\ast}_{u_3}(n\tdot p_2 )
+\,n_{u_1}n_{u_2}n^{\ast}_{u_4}n^{\ast}_{u_3}(n\tdot p_2 ) \;\;\;\;\; & & \mbox{}\nonumber \\
+\,n_{u_1}n_{u_2}n^{\ast}_{u_4}n^{\ast}_{u_3}(n\tdot p_3 )
+\,n_{u_1}n_{u_2}n^{\ast}_{u_4}n^{\ast}_{u_3}(n\tdot p_4 )\;\;\;\;\; & & \mbox{}\nonumber \\
+\,n_{u_1}n_{u_4}n^{\ast}_{u_2}n^{\ast}_{u_3}(n\tdot p_3 )
+\,n_{u_1}n_{u_4}n^{\ast}_{u_2}n^{\ast}_{u_3}(n\tdot p_4 ))\;\;\;\;\; & & \mbox{}\nonumber \\
+24\,(n_{u_1}n_{u_4}n^{\ast}_{u_3}n_{u_2}(n^{\ast}\tdot p_2 )
+\,n_{u_1}n_{u_4}n^{\ast}_{u_3}n_{u_2}(n^{\ast}\tdot p_3 )\;\;\;\;\; & & \mbox{}\nonumber \\
+\,n_{u_1}n_{u_4}n^{\ast}_{u_3}n_{u_2}(n^{\ast}\tdot p_4 ))\;\;\;\;\; & & \mbox{}\nonumber \\
+12\,(n_{u_1}n_{u_4}n^{\ast}_{u_2}n_{u_3}(n^{\ast}\tdot p_2 )
+\,n_{u_1}n_{u_2}n^{\ast}_{u_4}n_{u_3}(n^{\ast}\tdot p_2 )\;\;\;\;\; & & \mbox{}\nonumber \\
+\,n_{u_1}n_{u_2}n^{\ast}_{u_4}n_{u_3}(n^{\ast}\tdot p_3 )
+\,n_{u_1}n_{u_2}n^{\ast}_{u_4}n_{u_3}(n^{\ast}\tdot p_4 ))\;\;\;\;\; & & \mbox{}\nonumber \\
+\,n_{u_1}n_{u_4}n^{\ast}_{u_2}n_{u_3}(n^{\ast}\tdot p_3 )
+\,n_{u_1}n_{u_4}n^{\ast}_{u_2}n_{u_3}(n^{\ast}\tdot p_4 )\big]\;\;\;\;\; & & \mbox{}\nonumber \\
-\frac {1}{4\,(n\tdot n^{\ast})}
\big[ 2\,(n_{u_1}g_{u_2 u_3}n^{\ast}_{u_4}
+\,n_{u_1}g_{u_3 u_4}n^{\ast}_{u_2})\;\;\;\;\; & & \mbox{}\nonumber \\
+3\,(n_{u_2}n^{\ast}_{u_4}g_{u_1 u_3}
+\,n_{u_4}n^{\ast}_{u_3}g_{u_1 u_2}\;\;\;\;\; & & \mbox{}\nonumber \\
+\,n_{u_2}n^{\ast}_{u_3}g_{u_1 u_4}
+\,n_{u_4}n^{\ast}_{u_2}g_{u_1 u_3})\;\;\;\;\; & & \mbox{}\nonumber \\
+7\,(n_{u_2}g_{u_3 u_4}n^{\ast}_{u_1}
+\,g_{u_2 u_3}n_{u_4}n^{\ast}_{u_1})\;\;\;\;\; & & \mbox{}\nonumber \\
+ 30\,n_{u_1}g_{u_2 u_4}n^{\ast}_{u_3}\big] \bigg{\}}\;\;\;\;\; & & \mbox{}\nonumber \\
+(1\rightarrow 2,2\rightarrow 3,3\rightarrow 4, 4\rightarrow 1)\;\;\;\;\; & & \mbox{}\nonumber \\
+(1\rightarrow 3,2\rightarrow 4,3\rightarrow 1, 4\rightarrow 2)\;\;\;\;\; & & \mbox{}\nonumber \\
+(1\rightarrow 4,2\rightarrow 1,3\rightarrow 2, 4\rightarrow 3)\;\;\;\;\; & & \mbox{}\nonumber \\
%
%
+\overline{I} f^{a_1 ab}f^{a_2 bc} f^{a_3 cd}f^{a_4 da}
\bigg{\{}
\frac{1}{12}
\big[ 55\,g_{u_3 u_4}g_{u_1 u_2}
+g_{u_2 u_3}g_{u_1 u_4}\!\!\!\! & & \mbox{} \nonumber \\
+g_{u_2 u_4}g_{u_1 u_3}\big]\;\;\;\;\; & & \mbox{}\nonumber \\
-\frac{1}{(n\tdot n^{\ast})\, n\tdot (p_3 + p_4 )}
\big[ n_{u_1}n_{u_2}g_{u_3 u_4}(n^{\ast}\tdot p_3 )
+n_{u_1}n_{u_2}g_{u_3 u_4}(n^{\ast}\tdot p_4 )\;\;\;\;\; & & \mbox{}\nonumber \\
+n_{u_3}n_{u_4}g_{u_1 u_2}(n^{\ast}\tdot p_3 )
+n_{u_3}n_{u_4}g_{u_1 u_2}(n^{\ast}\tdot p_4 )\;\;\;\;\; & & \mbox{}\nonumber \\
+4\,(n_{u_1}n_{u_3}g_{u_2 u_4}(n^{\ast}\tdot p_3 )
+\,n_{u_1}n_{u_3}g_{u_2 u_4}(n^{\ast}\tdot p_4 )\;\;\;\;\; & & \mbox{}\nonumber \\
+\,n_{u_2}n_{u_4}g_{u_1 u_3}(n^{\ast}\tdot p_3 )
+\,n_{u_2}n_{u_4}g_{u_1 u_3}(n^{\ast}\tdot p_4 ))\big]\;\;\;\;\; & & \mbox{}\nonumber \\
-\frac {1}{4\,(n\tdot n^{\ast})^{2}\, n\tdot(p_3 + p_4 )}
\big[ 19(n\tdot n^{\ast})\,
(n_{u_4}n^{\ast}_{u_3}g_{u_1 u_2}(n\tdot p_3 )
+n_{u_4}n^{\ast}_{u_3}g_{u_1 u_2}(n\tdot p_4 ))\;\;\;\;\; & & \mbox{}\nonumber \\
-(n\tdot n^{\ast})(
n_{u_3}n^{\ast}_{u_1}g_{u_2 u_4}(n\tdot p_4 )
+n_{u_2}n^{\ast}_{u_3}g_{u_1 u_4}(n\tdot p_3 )\;\;\;\;\; & & \mbox{}\nonumber \\
+n_{u_2}n^{\ast}_{u_3}g_{u_1 u_4}(n\tdot p_4 )
+n_{u_2}n^{\ast}_{u_4}g_{u_1 u_3}(n\tdot p_3 )\;\;\;\;\; & & \mbox{}\nonumber \\
+n_{u_2}n^{\ast}_{u_4}g_{u_1 u_3}(n\tdot p_4 )
+n_{u_4}n^{\ast}_{u_2}g_{u_1 u_3}(n\tdot p_3 )\;\;\;\;\; & & \mbox{}\nonumber \\
+n_{u_4}n^{\ast}_{u_2}g_{u_1 u_3}(n\tdot p_4 )
+n_{u_1}g_{u_2 u_4}n^{\ast}_{u_3}(n\tdot p_3 )\;\;\;\;\; & & \mbox{}\nonumber \\
+n_{u_1}g_{u_2 u_4}n^{\ast}_{u_3}(n\tdot p_4 )
+n_{u_3}n^{\ast}_{u_2}g_{u_1 u_4}(n\tdot p_3 )\;\;\;\;\; & & \mbox{}\nonumber \\
+n_{u_3}n^{\ast}_{u_2}g_{u_1 u_4}(n\tdot p_4 )
+n_{u_3}n^{\ast}_{u_1}g_{u_2 u_4}(n\tdot p_3 )\;\;\;\;\; & & \mbox{}\nonumber \\
+n_{u_4}n^{\ast}_{u_1}g_{u_2 u_3}(n\tdot p_3 )
+n_{u_4}n^{\ast}_{u_1}g_{u_2 u_3}(n\tdot p_4 )\;\;\;\;\; & & \mbox{}\nonumber \\
+n_{u_1}g_{u_2 u_3}n^{\ast}_{u_4}(n\tdot p_3 )
+n_{u_1}g_{u_2 u_3}n^{\ast}_{u_4}(n\tdot p_4 ))\;\;\;\;\; & & \mbox{}\nonumber \\
+2\,(n_{u_2}n^{\ast}_{u_1}n^{\ast}_{u_4}n_{u_3}(n\tdot p_3 )
+\,n_{u_2}n^{\ast}_{u_1}n^{\ast}_{u_4}n_{u_3}(n\tdot p_4 )\;\;\;\;\; & & \mbox{}\nonumber \\
+\,n_{u_1}n_{u_4}n^{\ast}_{u_2}n^{\ast}_{u_3}(n\tdot p_3 )
+\,n_{u_1}n_{u_4}n^{\ast}_{u_2}n^{\ast}_{u_3}(n\tdot p_4 )\;\;\;\;\; & & \mbox{}\nonumber \\
+3\,n_{u_4}n^{\ast}_{u_2}n^{\ast}_{u_1}n_{u_3}(n\tdot p_3 )
+3\,n_{u_4}n^{\ast}_{u_2}n^{\ast}_{u_1}n_{u_3}(n\tdot p_4 )\;\;\;\;\; & & \mbox{}\nonumber \\
-7\,n_{u_1}n^{\ast}_{u_2}n^{\ast}_{u_4}n_{u_3}(n\tdot p_3 )
-7\,n_{u_1}n^{\ast}_{u_2}n^{\ast}_{u_4}n_{u_3}(n\tdot p_4 )\;\;\;\;\; & & \mbox{}\nonumber \\
-7\,n_{u_2}n^{\ast}_{u_1}n^{\ast}_{u_3}n_{u_4}(n\tdot p_3 )
-7\,n_{u_2}n^{\ast}_{u_1}n^{\ast}_{u_3}n_{u_4}(n\tdot p_4 ))\;\;\;\;\; & & \mbox{}\nonumber \\
-12\,(n_{u_1}n_{u_4}n^{\ast}_{u_3}n_{u_2}(n^{\ast}\tdot p_3 )
+\,n_{u_1}n_{u_4}n^{\ast}_{u_3}n_{u_2}(n^{\ast}\tdot p_4 )\;\;\;\;\; & & \mbox{}\nonumber\\
+\,n_{u_1}n_{u_4}n^{\ast}_{u_2}n_{u_3}(n^{\ast}\tdot p_3 )
+\,n_{u_1}n_{u_4}n^{\ast}_{u_2}n_{u_3}(n^{\ast}\tdot p_4 )\;\;\;\;\; & & \mbox{}\nonumber \\
+\,n_{u_1}n_{u_2}n^{\ast}_{u_4}n_{u_3}(n^{\ast}\tdot p_3 )
+\,n_{u_1}n_{u_2}n^{\ast}_{u_4}n_{u_3}(n^{\ast}\tdot p_4 ))\;\;\;\;\; & & \mbox{}\nonumber \\
+6\,(n_{u_1}n_{u_2}n^{\ast}_{u_4}n^{\ast}_{u_3}(n\tdot p_3 )
+\,n_{u_1}n_{u_2}n^{\ast}_{u_4}n^{\ast}_{u_3}(n\tdot p_4 ))\;\;\;\;\; & & \mbox{}\nonumber \\
+19\,(n\tdot n^{\ast})(n_{u_2}g_{u_3 u_4}n^{\ast}_{u_1}(n\tdot p_3 )
+\,n_{u_2}g_{u_3 u_4}n^{\ast}_{u_1}(n\tdot p_4 )\;\;\;\;\; & & \mbox{}\nonumber \\
+\,n_{u_1}g_{u_3 u_4}n^{\ast}_{u_2}(n\tdot p_3 )
+\,n_{u_1}g_{u_3 u_4}n^{\ast}_{u_2}(n\tdot p_4 )\;\;\;\;\; & & \mbox{}\nonumber \\
+\,n_{u_3}g_{u_1 u_2}n^{\ast}_{u_4}(n\tdot p_4 )
+\,n_{u_3}g_{u_1 u_2}n^{\ast}_{u_4}(n\tdot p_3 ))\;\;\;\;\; & & \mbox{}\nonumber \\
-12\,(n_{u_2}n_{u_4}n^{\ast}_{u_1}n_{u_3}(n^{\ast}\tdot p_3 )
+\,n_{u_2}n_{u_4}n^{\ast}_{u_1}n_{u_3}(n^{\ast}\tdot p_4 ))\big] \bigg{\}}\;\;\;\;\; & & \mbox{}\nonumber\\
+(1\rightarrow 2,2\rightarrow 3,3\rightarrow 4, 4\rightarrow 1).\;\;\;\;\; & & \mbox{}\nonumber
\end{eqnarray}

2). $J^{lynx}_{2}$, the integrated result of (\ref{eq:3p2lc}):
\begin{eqnarray}\label{eq:lx2lc}
J^{lynx}_{2}=
%
%
\frac {-\overline{I}f^{a_1 ab}f^{a_2 bc} f^{a_3 cd}f^{a_4 da}}{2 n\tdot (p_2 +p_3 )(n\tdot n^{\ast})^{2}(n\tdot p_3 )(n\tdot p_2 )}
\,\bigg{\{} \;\;\;\;\;\;& &  \\
(n\tdot n^{\ast})^{2} \big[ 
(n\tdot p_2 )^{2}g_{u_3 u_4}g_{u_1 u_2}(n\tdot p_3 )
+(n\tdot p_3 )^{2}g_{u_3 u_4}g_{u_1 u_2}(n\tdot p_2 )\;\;\;\;\; & & \mbox{} \nonumber \\
+(n\tdot p_3 )^{2}g_{u_1 u_3}g_{u_2 u_4}(n\tdot p_2 ) 
+(n\tdot p_2 )^{2}g_{u_1 u_3}g_{u_2 u_4}(n\tdot p_3 )
\big]\;\;\;\;\; & & \mbox{}\nonumber \\
+2\big[\,(n\tdot p_3 )^{2}n_{u_4}n_{u_2}(n^{\ast}\tdot p_2 )n_{u_3}n^{\ast}_{u_1}
+\,(n\tdot p_2 )^{2}n_{u_3}n_{u_4}n^{\ast}_{u_1}n_{u_2}(n^{\ast}\tdot p_3 )\;\;\;\;\; & & \mbox{}\nonumber \\
+\,(n\tdot p_2 )^{2}n_{u_3}n_{u_2}n^{\ast}_{u_4}n_{u_1}(n^{\ast}\tdot p_3 )
+\,(n\tdot p_3 )^{2}n_{u_1}n_{u_2}(n^{\ast}\tdot p_2 )n_{u_3}n^{\ast}_{u_4}
\big]\;\;\;\;\; & & \mbox{}\nonumber \\
-2(n\tdot n^{\ast})\big[\,
(n\tdot p_2 )^{2}n_{u_3}n_{u_1}g_{u_2 u_4}(n^{\ast}\tdot p_3 )
+\,(n\tdot p_2 )^{2}n_{u_3}n_{u_4}g_{u_1 u_2}(n^{\ast}\tdot p_3 )\;\;\;\;\; & & \mbox{}\nonumber \\
+\,(n\tdot p_3 )^{2}n_{u_4}n_{u_2}g_{u_1 u_3}(n^{\ast}\tdot p_2 )
+\,(n\tdot p_3 )^{2}g_{u_3 u_4}n_{u_1}n_{u_2}(n^{\ast}\tdot p_2 )\;\;\;\;\; & & \mbox{}\nonumber \\
+\,n_{u_4}(n\tdot p_2 )^{2}n^{\ast}_{u_3}g_{u_1 u_2}(n\tdot p_3 )
+\,n_{u_1}(n\tdot p_2 )^{2}n^{\ast}_{u_3}g_{u_2 u_4}(n\tdot p_3 )\;\;\;\;\; & & \mbox{}\nonumber \\
+\,n_{u_4}n^{\ast}_{u_3}g_{u_1 u_2}(n\tdot p_3 )^{2}(n\tdot p_2 )
+\,n_{u_1}n^{\ast}_{u_2}g_{u_3 u_4}(n\tdot p_3 )^{2}(n\tdot p_2 )\;\;\;\;\; & & \mbox{}\nonumber \\
+\,n_{u_4}n^{\ast}_{u_2}g_{u_1 u_3}(n\tdot p_3 )^{2}(n\tdot p_2 )
+\,n_{u_1}n^{\ast}_{u_3}g_{u_2 u_4}(n\tdot p_3 )^{2}(n\tdot p_2 )\;\;\;\;\; & & \mbox{}\nonumber \\
+\,g_{u_3 u_4}(n\tdot p_2 )^{2}n_{u_1}n^{\ast}_{u_2}(n\tdot p_3 )
+\,g_{u_1 u_3}(n\tdot p_2 )^{2}n_{u_4}n^{\ast}_{u_2}(n\tdot p_3 )
\big]\;\;\;\;\; & & \mbox{}\nonumber \\
+4(n\tdot n^{\ast})(n\tdot p_2 )(n\tdot p_3 )\big[
\,g_{u_2 u_3}n_{u_4}n_{u_1}(n^{\ast}\tdot p_2 )
+\,g_{u_2 u_3}n_{u_4}n_{u_1}(n^{\ast}\tdot p_3 )\;\;\;\;\; & & \mbox{}\nonumber \\
-\,n_{u_2}n_{u_4}g_{u_1 u_3}(n^{\ast}\tdot p_3 )
-\,n_{u_1}n_{u_2}g_{u_3 u_4}(n^{\ast}\tdot p_3 )\;\;\;\;\; & & \mbox{}\nonumber \\
-\,n_{u_3}n_{u_4}g_{u_1 u_2}(n^{\ast}\tdot p_2 )
-\,n_{u_3}n_{u_1}g_{u_2 u_4}(n^{\ast}\tdot p_2 )
\big]\;\;\;\;\; & & \mbox{}\nonumber \\
+6(n\tdot p_2 )(n\tdot p_3 )\big[
\,n_{u_3}n_{u_4}n^{\ast}_{u_1}n_{u_2}(n^{\ast}\tdot p_3)
+\,n_{u_3}n_{u_2}n^{\ast}_{u_4}n_{u_1}(n^{\ast}\tdot p_3)\;\;\;\;\; & & \mbox{}\nonumber \\
+\,n_{u_4}n_{u_2}(n^{\ast}\tdot p_2 )n_{u_3}n^{\ast}_{u_1}
+\,n_{u_1}n_{u_2}(n^{\ast}\tdot p_2 )n_{u_3}n^{\ast}_{u_4}\;\;\;\;\; & & \mbox{}\nonumber \\
-\,n_{u_4}n_{u_2}g_{u_1 u_3}(n^{\ast}\tdot p_2 )(n\tdot n^{\ast})
-\,n_{u_3}n_{u_1}g_{u_2 u_4}(n^{\ast}\tdot p_3 )(n\tdot n^{\ast})\;\;\;\;\; & & \mbox{}\nonumber \\
-\,g_{u_3 u_4}n_{u_1}n_{u_2}(n^{\ast}\tdot p_2 )(n\tdot n^{\ast})
-\,n_{u_3}n_{u_4}g_{u_1 u_2}(n^{\ast}\tdot p_3 )(n\tdot n^{\ast})
\big]\;\;\;\;\; & & \mbox{}\nonumber \\
+8(n\tdot p_2 )(n\tdot p_3 ) \big[
\,n_{u_3}(n\tdot p_2 )n^{\ast}_{u_2}n^{\ast}_{u_4}n_{u_1}
+\,n_{u_2}n_{u_1}n^{\ast}_{u_4}n^{\ast}_{u_3}(n\tdot p_3 )\;\;\;\;\; & & \mbox{}\nonumber \\
+\,n_{u_3}n_{u_1}n^{\ast}_{u_2}n^{\ast}_{u_4}(n\tdot p_3 )
+\,n_{u_4}n^{\ast}_{u_1}n^{\ast}_{u_2}n_{u_3}(n\tdot p_3 )\;\;\;\;\; & & \mbox{}\nonumber \\
+\,n_{u_4}n^{\ast}_{u_1}n^{\ast}_{u_3}n_{u_2}(n\tdot p_3 )
+\,n_{u_1}n_{u_2}n^{\ast}_{u_3}n^{\ast}_{u_4}(n\tdot p_2 )\;\;\;\;\; & & \mbox{}\nonumber \\
+\,n_{u_4}n_{u_2}n^{\ast}_{u_3}n^{\ast}_{u_1}(n\tdot p_2 )
+\,n_{u_4}(n\tdot p_2 )n^{\ast}_{u_1}n^{\ast}_{u_2}n_{u_3}\;\;\;\;\; & & \mbox{}\nonumber \\
-\,n_{u_4}n^{\ast}_{u_1}g_{u_2 u_3}(n\tdot p_3 )(n\tdot n^{\ast})
-\,n_{u_1}n^{\ast}_{u_4}g_{u_2 u_3}(n\tdot p_3 )(n\tdot n^{\ast})\;\;\;\;\; & & \mbox{}\nonumber \\
-\,n_{u_1}(n\tdot p_2 )n^{\ast}_{u_4}g_{u_2 u_3}(n\tdot n^{\ast})
-\,n_{u_4}(n\tdot p_2 )n^{\ast}_{u_1}g_{u_2 u_3}(n\tdot n^{\ast})
\big]\;\;\;\;\; & & \mbox{}\nonumber \\
+12(n\tdot p_2 )(n\tdot p_3 )\big[
\,n_{u_3}n_{u_4}n^{\ast}_{u_2}n_{u_1}(n^{\ast}\tdot p_2)
+\,n_{u_2}n_{u_4}n^{\ast}_{u_3}n_{u_1}(n^{\ast}\tdot p_3 )\;\;\;\;\; & & \mbox{}\nonumber \\
+\,n_{u_4}n_{u_1}n^{\ast}_{u_2}n^{\ast}_{u_3}(n\tdot p_3 )
+\,n_{u_4}(n\tdot p_2 )n^{\ast}_{u_3}n^{\ast}_{u_2}n_{u_1}
\big]\;\;\;\;\; & & \mbox{}\nonumber \\
+12\big[
\,(n\tdot p_2 )^{2}n_{u_3}n_{u_4}n^{\ast}_{u_2}n_{u_1}(n^{\ast}\tdot p_3 )
+\,(n\tdot p_3 )^{2}n_{u_1}n_{u_2}(n^{\ast}\tdot p_2 )n_{u_4}n^{\ast}_{u_3}\;\;\;\;\; & & \mbox{}\nonumber \\
-\,(n\tdot p_3 )^{2}g_{u_1 u_4}n_{u_2}n_{u_3}(n^{\ast}\tdot p_2 )(n\tdot n^{\ast})
-\,(n\tdot p_2 )^{2}g_{u_1 u_4}n_{u_2}n_{u_3}(n^{\ast}\tdot p_3 )(n\tdot n^{\ast})
\big]\;\;\;\;\; & & \mbox{}\nonumber \\
-16(n\tdot n^{\ast})(n\tdot p_2)(n\tdot p_3) \big[
\,g_{u_1 u_4}n_{u_2}n_{u_3}(n^{\ast}\tdot p_2 )
+\,g_{u_1 u_4}n_{u_2}n_{u_3}(n^{\ast}\tdot p_3 )
\big]\;\;\;\;\; & & \mbox{}\nonumber \\
+17(n\tdot n^{\ast})^{2}(n\tdot p_2)(n\tdot p_3)\big[
\,g_{u_1 u_4}(n\tdot p_2 )g_{u_2 u_3}
+\,g_{u_2 u_3}g_{u_1 u_4}(n\tdot p_3 )
\big]\;\;\;\;\; & & \mbox{}\nonumber \\
+24(n\tdot p_2 )(n\tdot p_3)\big[
\,n_{u_3}n_{u_4}n^{\ast}_{u_2}n_{u_1}(n^{\ast}\tdot p_3 )
+\,n_{u_1}n_{u_2}(n^{\ast}\tdot p_2 )n_{u_4}n^{\ast}_{u_3}\;\;\;\;\; & & \mbox{}\nonumber \\
-\,n_{u_3}n^{\ast}_{u_1}n^{\ast}_{u_4}n_{u_2}(n\tdot p_3 )
-\,n_{u_3}n_{u_2}n^{\ast}_{u_1}n^{\ast}_{u_4}(n\tdot p_2 )
\big]\;\;\;\;\; & & \mbox{}\nonumber \\
-30(n\tdot n^{\ast})(n\tdot p_2 )(n\tdot p_3 )\big[
\,n_{u_2}n^{\ast}_{u_3}g_{u_1 u_4}(n\tdot p_3 )
+\,n_{u_3}n^{\ast}_{u_2}g_{u_1 u_4}(n\tdot p_3 )\;\;\;\;\; & & \mbox{}\nonumber \\
+\,g_{u_1 u_4}(n\tdot p_2 )n_{u_2}n^{\ast}_{u_3}
+\,g_{u_1 u_4}(n\tdot p_2 )n_{u_3}n^{\ast}_{u_2}
\big]
\bigg{\}}\;\;\;\;\; & & \mbox{}\nonumber\\
%
%
+\frac{N}{2} f^{a_1 a_4 e}f^{a_2 a_3 e}
\frac {\overline{I}}{4n\tdot (p_2 +p_3 )
(n\tdot n^{\ast})^{2}(n\tdot p_2 )(n\tdot p_3 )}\;
\bigg{\{} \;\;\;\;\; & & \mbox \nonumber \\ 
2\,\big[
\,n_{u_4}n^{\ast}_{u_1}n^{\ast}_{u_3}n_{u_2}(n\tdot p_3 )^{2}(n\tdot p_2)
+\,n_{u_4}n_{u_2}n^{\ast}_{u_3}n^{\ast}_{u_1}(n\tdot p_2 )^{2}(n\tdot p_3 )\;\;\;\;\; & & \mbox{}\nonumber \\
+\,n_{u_3}(n\tdot p_2 )^{2}n^{\ast}_{u_2}n^{\ast}_{u_4}n_{u_1}(n\tdot p_3 )
+\,n_{u3}n_{u_1}n^{\ast}_{u_2}n^{\ast}_{u_4}(n\tdot p_3 )^{2}(n\tdot p_2 )\;\;\;\;\; & & \mbox{}\nonumber \\%
-(n\tdot n^{\ast})^{2}\,((n\tdot p_3 )^{2}g_{u_1 u_3}g_{u_2 u_4}(n\tdot p_2 )
+\,(n\tdot p_2 )^{2}g_{u_1 u_3}g_{u_2 u_4}(n\tdot p_3 ))\big]\;\;\;\;\; & & \mbox{}\nonumber \\
+4\,\big[(n\tdot n^{\ast})(n\tdot p_2 )(n\tdot p_3 )(
\,g_{u_2 u_3}n_{u_4}n_{u_1}(n^{\ast}\tdot p_2 )
+\,g_{u_2 u_3}n_{u_4}n_{u_1}(n^{\ast}\tdot p_3 )\;\;\;\;\; & & \mbox{}\nonumber \\
+\,g_{u_3 u_4}(n\tdot p_2 )n_{u_1}n^{\ast}_{u_2}
+\,n_{u_4}(n\tdot p_2 )n^{\ast}_{u_3}g_{u_1 u_2}\;\;\;\;\; & & \mbox{}\nonumber \\
+\,n_{u_4}n^{\ast}_{u_3}g_{u_1 u_2}(n\tdot p_3 )
+\,n_{u_1}n^{\ast}_{u_2}g_{u_3 u_4}(n\tdot p_3 ))\;\;\;\;\; & & \mbox{}\nonumber \\
+(n\tdot n^{\ast})^{2}(n\tdot p_2 )(n\tdot p_3 )(
(n\tdot p_2 )g_{u_3 u_4}g_{u_1 u_2}
+\,(n\tdot p_3 )g_{u_3 u_4}g_{u_1,u_2})\;\;\;\;\; & & \mbox{}\nonumber \\
-\,(n\tdot p_2 )^{2}n_{u_3}n_{u_4}n^{\ast}_{u_1}n_{u_2}(n^{\ast}\tdot p_3 )
-\,(n\tdot p_3 )^{2}n_{u_1}n_{u_2}(n^{\ast}\tdot p_2 )n_{u_3}n^{\ast}_{u_4}
\big]\;\;\;\;\; & & \mbox{}\nonumber \\
+8\,\big[
(n\tdot p_2 )^{2}n_{u_3}n_{u_2}n^{\ast}_{u_4}n_{u_1}(n^{\ast}\tdot p_3 )
+\,(n\tdot p_3 )^{2}n_{u_4}n_{u_2}( n^{\ast}\tdot p_2 )n_{u_3}n^{\ast}_{u_1}\;\;\;\;\; & & \mbox{}\nonumber \\
+(n\tdot n^{\ast})(n\tdot p_2)(n\tdot p_3 )
(n_{u_2}n_{u_4}g_{u_1 u_3}(n^{\ast}\tdot p_3 )
+\,n_{u_3}n_{u_1}g_{u_2 u_4}(n^{\ast}\tdot p_2 )\;\;\;\;\; & & \mbox{}\nonumber \\
-\,n_{u_4}(n\tdot p_2 )n^{\ast}_{u_2}g_{u_1 u_3}
-\,n_{u_1}(n\tdot p_2 )n^{\ast}_{u_3}g_{u_2 u_4}\;\;\;\;\; & & \mbox{}\nonumber \\
-\,n_{u_1}(n\tdot p_2 )n^{\ast}_{u_4}g_{u_2 u_3}
-\,n_{u_4}(n\tdot p_2 )n^{\ast}_{u_1}g_{u_2 u_3}\;\;\;\;\; & & \mbox{}\nonumber \\
-\,n_{u_1}n^{\ast}_{u_4}g_{u_2 u_3}(n\tdot p_3 )
-\,n_{u_4}n^{\ast}_{u_1}g_{u_2 u_3}(n\tdot p_3 )\;\;\;\;\; & & \mbox{}\nonumber \\
-\,n_{u_4}n^{\ast}_{u_2}g_{u_1 u_3}(n\tdot p_3 )
-\,n_{u_1}n^{\ast}_{u_3}g_{u_2 u_4}(n\tdot p_3 ))
\big]\;\;\;\;\; & & \mbox{}\nonumber \\
+12\,\big[n_{u_3}n_{u_4}n^{\ast}_{u_2}n_{u_1}(n^{\ast}\tdot p_2 )(n\tdot p_3 )(n\tdot p_2 )
+\,n_{u_4}n_{u_2}(n^{\ast}\tdot p_2 )n_{u_3}n^{\ast}_{u_1}(n\tdot p_3 )(n\tdot p_2 )\;\;\;\;\; & & \mbox{}\nonumber \\
+\,n_{u_3}n_{u_2}n^{\ast}_{u_4}n_{u_1}(n^{\ast}\tdot p_3 )(n\tdot p_3 )(n\tdot p_2 )
+\,n_{u_4}n_{u_1}n^{\ast}_{u_2}n^{\ast}_{u_3}(n\tdot p_3 )^{2}(n\tdot p_2 )\;\;\;\;\; & & \mbox{}\nonumber \\
+\,n_{u_2}n_{u_4}n^{\ast}_{u_3}n_{u_1}(n^{\ast}\tdot p_3 )(n\tdot p_3 )(n\tdot p_2 )
+\,n_{u_4}(n\tdot p_2 )^{2}n^{\ast}_{u_3}n^{\ast}_{u_2}n_{u_1}(n\tdot p_3 )\;\;\;\;\; & & \mbox{}\nonumber \\
+\,(n\tdot p_2 )^{2}n_{u_3}n_{u_4}n^{\ast}_{u_2}n_{u_1}(n^{\ast}\tdot p_3 )
+\,(n\tdot p_3 )^{2}n_{u_1}n_{u_2}(n^{\ast}\tdot p_2 )n_{u_4}n^{\ast}_{u_3}\;\;\;\;\; & & \mbox{}\nonumber \\
-(n\tdot n^{\ast})
(\,n_{u_3}n_{u_1}g_{u_2 u_4}(n^{\ast}\tdot p_3 )(n\tdot p_3 )(n\tdot p_2 )
+\,n_{u_4}n_{u_2}g_{u_1 u_3}( n^{\ast}\tdot p_2 )(n\tdot p_3 )(n\tdot p_2 )\;\;\;\;\; & & \mbox{}\nonumber \\
+\,(n\tdot p_3 )^{2}g_{u_1 u_4}n_{u_2}n_{u_3}(n^{\ast}\tdot p_2 )
+\,(n\tdot p_2 )^{2}g_{u_1 u_4}n_{u_2}n_{u_3}(n^{\ast}\tdot p_3 ))
\big] \;\;\;\;\; & & \mbox{}\nonumber \\
+14(n\tdot p_2 )(n\tdot p_3) \big[n_{u_2}n_{u_1}n^{\ast}_{u_4}n^{\ast}_{u_3}(n\tdot p_3 )
+\,n_{u_1}n_{u_2}n^{\ast}_{u_3}n^{\ast}_{u_4}(n\tdot p_2 )\;\;\;\;\; & & \mbox{}\nonumber \\
+\,n_{u_4}n^{\ast}_{u_1}n^{\ast}_{u_2}n_{u_3}(n\tdot p_3 )
+\,n_{u_4}(n\tdot p_2 )n^{\ast}_{u_1}n^{\ast}_{u_2}n_{u_3}
\big]\;\;\;\;\; & & \mbox{}\nonumber \\
+16\,(n\tdot n^{\ast})\big[
(n\tdot p_2 )^{2}n_{u_3}n_{u_4}g_{u_1 u_2}(n^{\ast}\tdot p_3 )
+\,(n\tdot p_3 )^{2}g_{u_3 u_4}n_{u_1}n_{u_2}(n^{\ast}\tdot p_2 )\;\;\;\;\; & & \mbox{}\nonumber \\
-(n\tdot p_3 )(n\tdot p_2 )(\,n_{u_3}n_{u_4}g_{u_1 u_2}(n^{\ast}\tdot p_2 )
+\,n_{u_1}n_{u_2}g_{u_3 u_4}(n^{\ast}\tdot p_3 )\;\;\;\;\; & & \mbox{}\nonumber \\
+\,g_{u_1 u_4}n_{u_2}n_{u_3}(n^{\ast}\tdot p_2 )
+\,g_{u_1 u_4}n_{u_2}n_{u_3}(n^{\ast}\tdot p_3 ))
\big]\;\;\;\;\; & & \mbox{}\nonumber \\
+17\,(n\tdot n^{\ast})^{2}(n\tdot p_2)(n\tdot p_3)\big[
g_{u_1 u_4}(n\tdot p_2 )g_{u_2 u_3}
+\,g_{u_2 u_3}g_{u_1 u_4}(n\tdot p_3 )
\big]\;\;\;\;\; & & \mbox{}\nonumber \\
+18\,(n\tdot n^{\ast})(n\tdot p_2)(n\tdot p_3)\big[
(n\tdot p_3 )n_{u_2}n^{\ast}_{u_4}g_{u_1 u_3}
+\,(n\tdot p_2 )n_{u_3}n^{\ast}_{u_1}g_{u_2 u_4}\;\;\;\;\; & & \mbox{}\nonumber \\
+\,(n\tdot p_3 )n_{u_3}n^{\ast}_{u_1}g_{u_2 u_4}
+\,(n\tdot p_2 )g_{u_1 u_3}n_{u_2}n^{\ast}_{u_4}\;\;\;\;\; & & \mbox{}\nonumber \\
-\,(n\tdot p_3 )n_{u_3}n^{\ast}_{u_4}g_{u_1 u_2}
-\,(n\tdot p_2 )g_{u_3 u_4}n_{u_2}n^{\ast}_{u_1}\;\;\;\;\; & & \mbox{}\nonumber \\
-\,(n\tdot p_2 )n_{u_3}n^{\ast}_{u_4}g_{u_1 u_2}
-\,(n\tdot p_3 )n_{u_2}n^{\ast}_{u_1}g_{u_3 u_4}
\big]\;\;\;\;\; & & \mbox{}\nonumber \\
-20\,(n\tdot n^{\ast})\big[
(n\tdot p_2 )^{2}n_{u_3}n_{u_1}g_{u_2 u_4}(n^{\ast}\tdot p_3 )
+\,(n\tdot p_3 )^{2}n_{u_4}n_{u_2}g_{u_1 u_3}(n^{\ast}\tdot p_2 )
\big]\;\;\;\;\; & & \mbox{}\nonumber \\
+24\,(n\tdot p_2 )(n\tdot p_3 )\big[n_{u_3}n_{u_4}n^{\ast}_{u_2}n_{u_1}(n^{\ast}\tdot p_3 )
+\,n_{u_1}n_{u_2}(n^{\ast}\tdot p_2 )n_{u_4}n^{\ast}_{u_3}\;\;\;\;\; & & \mbox{}\nonumber \\
-\,n_{u_3}n^{\ast}_{u_1}n^{\ast}_{u_4}n_{u_2}(n\tdot p_3 )
-\,n_{u_3}n_{u_2}n^{\ast}_{u_1}n^{\ast}_{u_4}(n\tdot p_2 )
\big]\;\;\;\;\; & & \mbox{}\nonumber \\
-30\,(n\tdot n^{\ast})g_{u_1 u_4}
\big[n_{u_3}n^{\ast}_{u_2}(n\tdot p_3 )^{2}(n\tdot p_2 )
+\,n_{u_2}n^{\ast}_{u_3}(n\tdot p_3 )^{2}(n\tdot p_2 )\;\;\;\;\; & & \mbox{}\nonumber \\
+\,n_{u_2}n^{\ast}_{u_3}(n\tdot p_2 )^{2}(n\tdot p_3 )
+\,n_{u_3}n^{\ast}_{u_2}(n\tdot p_2 )^{2}(n\tdot p_3 )\big]
\bigg{\}}\;\;\;\;\; & & \mbox{}\nonumber 
\end{eqnarray}

3.) $J^{fish}_{2}$, the integrated result of (\ref{eq:2p2lc}):
\begin{eqnarray} \label{eq:fsh2lc}
J^{fish}_{2}=
\frac {\overline{I}}{(n\tdot n^{\ast})^{2}n\tdot (p_2 +p_3)}
f^{a_1 ab}f^{a_2 bc} f^{a_3 cd}f^{a_4 da}
\bigg{\{}4\,(n\tdot n^{\ast})^{2} \,\big[g_{u_2 u_3}g_{u_1 u_4}(n\tdot p_2)\;\;\;\;\; & & \mbox{}  \\
+\,g_{u_2 u_3}g_{u_1 u_4}(n\tdot p_3)\big] \;\;\;\;\; & & \mbox{} \nonumber \\
+n_{u_2}n_{u_4}n^{\ast}_{u_1}n^{\ast}_{u_3}(n\tdot p_2)
+n_{u_2}n_{u_4}n^{\ast}_{u_1}n^{\ast}_{u_3}(n\tdot p_3)\;\;\;\;\; & & \mbox{} \nonumber \\
+n_{u_3}n_{u_1}n^{\ast}_{u_4}n^{\ast}_{u_2}(n\tdot p_3)
+n_{u_1}n_{u_2}n^{\ast}_{u_3}n^{\ast}_{u_4}(n\tdot p_2)\;\;\;\;\; & & \mbox{} \nonumber \\
+n_{u_1}n_{u_2}n^{\ast}_{u_3}n^{\ast}_{u_4}(n\tdot p_3)
+n_{u_3}n_{u_4}n^{\ast}_{u_1}n^{\ast}_{u_2}(n\tdot p_2)\;\;\;\;\; & & \mbox{} \nonumber \\
+n_{u_3}n_{u_4}n^{\ast}_{u_1}n^{\ast}_{u_2}(n\tdot p_3)
+n_{u_3}n_{u_1}n^{\ast}_{u_4}n^{\ast}_{u_2}(n\tdot p_2)\;\;\;\;\; & & \mbox{} \nonumber \\
+2\,\big[ n_{u_1}n_{u_4}n^{\ast}_{u_3}n^{\ast}_{u_2}(n\tdot p_2)
+\,n_{u_1}n_{u_4}n^{\ast}_{u_3}n^{\ast}_{u_2}(n\tdot p_3)\;\;\;\;\; & & \mbox{} \nonumber \\
+\,n_{u_3}n_{u_2}n^{\ast}_{u_1}n^{\ast}_{u_4}(n\tdot p_2)
+\,n_{u_3}n_{u_2}n^{\ast}_{u_1}n^{\ast}_{u_4}(n\tdot p_3)\big]\;\;\;\;\; & & \mbox{} \nonumber \\
+4\,\big[ n_{u_3}n_{u_4}n_{u_2}n^{\ast}_{u_1}(n^{\ast}\tdot p_2)
+\,n_{u_3}n_{u_4}n_{u_2}n^{\ast}_{u_1}(n^{\ast}\tdot p_3)\;\;\;\;\; & & \mbox{} \nonumber \\
+\,n_{u_3}n_{u_2}n_{u_1}n^{\ast}_{u_4}(n^{\ast}\tdot p_2)
+\,n_{u_3}n_{u_2}n_{u_1}n^{\ast}_{u_4}(n^{\ast}\tdot p_3)\;\;\;\;\; & & \mbox{} \nonumber \\
+\,n_{u_3}n_{u_4}n_{u_1}n^{\ast}_{u_2}(n^{\ast}\tdot p_2)
+\,n_{u_3}n_{u_4}n_{u_1}n^{\ast}_{u_2}(n^{\ast}\tdot p_3)\;\;\;\;\; & & \mbox{} \nonumber \\
+\,n_{u_2}n_{u_4}n_{u_1}n^{\ast}_{u_3}(n^{\ast}\tdot p_2)
+\,n_{u_2}n_{u_4}n_{u_1}n^{\ast}_{u_3}(n^{\ast}\tdot p_3)\big]\;\;\;\;\; & & \mbox{} \nonumber \\
+(n\tdot n^{\ast})
\big[2(g_{u_1 u_4}n_{u_3}n_{u_2}(n^{\ast}\tdot p_2)
+g_{u_1 u_4}n_{u_3}n_{u_2}(n^{\ast}\tdot p_3))
\;\;\;\;\; & & \mbox{} \nonumber \\
-n_{u_3}n_{u_1}g_{u_2 u_4}(n^{\ast}\tdot p_2)
-n_{u_1}n_{u_2}g_{u_3 u_4}(n^{\ast}\tdot p_2)\;\;\;\;\; & & \mbox{} \nonumber \\
-n_{u_3}n_{u_4}g_{u_1 u_2}(n^{\ast}\tdot p_3)
-n_{u_3}n_{u_1}g_{u_2 u_4}(n^{\ast}\tdot p_3)\;\;\;\;\; & & \mbox{} \nonumber \\
-n_{u_2}n_{u_4}g_{u_1 u_3}(n^{\ast}\tdot p_3)
-n_{u_1}n_{u_2}g_{u_3 u_4}(n^{\ast}\tdot p_3)\;\;\;\;\; & & \mbox{} \nonumber \\
-n_{u_3}n_{u_4}g_{u_1 u_2}(n^{\ast}\tdot p_2)
-n_{u_2}n_{u_4}g_{u_1 u_3}(n^{\ast}\tdot p_2)\;\;\;\;\; & & \mbox{} \nonumber \\
+2\,(g_{u_2 u_3}n_{u_4}n_{u_1}(n^{\ast}\tdot p_2)
+\,g_{u_2 u_3}n_{u_4}n_{u_1}(n^{\ast}\tdot p_3)\;\;\;\;\; & & \mbox{} \nonumber \\
-\,g_{u_1 u_4}n_{u_2}n^{\ast}_{u_3}(n\tdot p_2)
-\,g_{u_2 u_3}n_{u_4}n^{\ast}_{u_1}(n\tdot p_3)\;\;\;\;\; & & \mbox{} \nonumber \\
-\,g_{u_2 u_3}n_{u_4}n^{\ast}_{u_1}(n\tdot p_2)
-\,g_{u_2 u_3}n_{u_1}n^{\ast}_{u_4}(n\tdot p_3)\;\;\;\;\; & & \mbox{} \nonumber \\
-\,g_{u_1 u_4}n_{u_2}n^{\ast}_{u_3}(n\tdot p_3)
-\,g_{u_1 u_4}n_{u_3}n^{\ast}_{u_2}(n\tdot p_2)\;\;\;\;\; & & \mbox{} \nonumber \\
-\,g_{u_1 u_4}n_{u_3}n^{\ast}_{u_2}(n\tdot p_3)
-\,g_{u_2 u_3}n_{u_1}n^{\ast}_{u_4}(n\tdot p_2))
\big]
\bigg{\}}\;\;\;\;\; & & \mbox{} \nonumber \\
+\frac {\overline{I}}{(n\tdot n^{\ast})^{2}n\tdot (p_2 +p_3 )}
\frac{N}{2} f^{a_1 a_4 e}f^{a_2 a_3 e}
\bigg{\{}
-2(n\tdot n^{\ast})^2 \big[\,g_{u_2 u_3}g_{u_1 u_4}(n\tdot p_2)
+\,g_{u_2 u_3}g_{u_1 u_4}(n\tdot p_3)\big]\nonumber\\
+4\,\big[n_{u_2}n_{u_4}n^{\ast}_{u_1}n^{\ast}_{u_3}(n\tdot p_2)
+\,n_{u_2}n_{u_4}n^{\ast}_{u_1}n^{\ast}_{u_3}(n\tdot p_3 )\;\;\;\;\; & & \mbox{} \nonumber \\
+\,n_{u_3}n_{u_1}n^{\ast}_{u_4}n^{\ast}_{u_2}(n\tdot p_3 )
+\,n_{u_3}n_{u_1}n^{\ast}_{u_4}n^{\ast}_{u_2}(n\tdot p_2 )\big]\;\;\;\;\; & & \mbox{} \nonumber \\
-5\,\big[n_{u_1}n_{u_2}n^{\ast}_{u_3}n^{\ast}_{u_4}(n\tdot p_3 )
+\,n_{u_3}n_{u_4}n^{\ast}_{u_1}n^{\ast}_{u_2}(n\tdot p_2 )\;\;\;\;\; & & \mbox{} \nonumber \\
+\,n_{u_3}n_{u_4}n^{\ast}_{u_1}n^{\ast}_{u_2}(n\tdot p_3 )
+\,n_{u_1}n_{u_2}n^{\ast}_{u_3}n^{\ast}_{u_4}(n\tdot p_2 )\big]\;\;\;\;\; & & \mbox{} \nonumber \\
-n_{u_1}n_{u_4}n^{\ast}_{u_3}n^{\ast}_{u_2}(n\tdot p_2 )
-n_{u_1}n_{u_4}n^{\ast}_{u_3}n^{\ast}_{u_2}(n\tdot p_3 )\;\;\;\;\; & & \mbox{} \nonumber \\
-n_{u_3}n_{u_2}n^{\ast}_{u_1}n^{\ast}_{u_4}(n\tdot p_2)
-n_{u_3}n_{u_2}n^{\ast}_{u_1}n^{\ast}_{u_4}(n\tdot p_3 )\;\;\;\;\; & & \mbox{} \nonumber \\
-2\,\big[n_{u_3}n_{u_4}n_{u_2}n^{\ast}_{u_1}(n^{\ast}\tdot p_2 )
+\,n_{u_3}n_{u_4}n_{u_2}n^{\ast}_{u_1}(n^{\ast}\tdot p_3 )\;\;\;\;\; & & \mbox{} \nonumber \\
+\,n_{u_3}n_{u_2}n_{u_1}n^{\ast}_{u_4}(n^{\ast}\tdot p_2 )
+\,n_{u_3}n_{u_2}n_{u_1}n^{\ast}_{u_4}(n^{\ast}\tdot p_3 )\;\;\;\;\; & & \mbox{} \nonumber \\
+\,n_{u_3}n_{u_4}n_{u_1}n^{\ast}_{u_2}(n^{\ast}\tdot p_2)
+\,n_{u_3}n_{u_4}n_{u_1}n^{\ast}_{u_2}(n^{\ast}\tdot p_3 )\;\;\;\;\; & & \mbox{} \nonumber \\
+\,n_{u_2}n_{u_4}n_{u_1}n^{\ast}_{u_3}(n^{\ast}\tdot p_2)
+\,n_{u_2}n_{u_4}n_{u_1}n^{\ast}_{u_3}(n^{\ast}\tdot p_3 )\big]\;\;\;\;\; & & \mbox{} \nonumber \\
+(n\tdot n^{\ast})\,\big[g_{u_1 u_4}n_{u_2}n^{\ast}_{u_3}(n\tdot p_2 )
+g_{u_2 u_3}n_{u_4}n^{\ast}_{u_1}(n\tdot p_3 )\;\;\;\;\; & & \mbox{} \nonumber \\
+g_{u_2 u_3}n_{u_1}n^{\ast}_{u_4}(n\tdot p_3 )
+g_{u_1 u_4}n_{u_2}n^{\ast}_{u_3}(n\tdot p_3 )\;\;\;\;\; & & \mbox{} \nonumber \\
+g_{u_1 u_4}n_{u_3}n^{\ast}_{u_2}(n\tdot p_2 )
+g_{u_1 u_4}n_{u_3}n^{\ast}_{u_2}(n\tdot p_3 )\;\;\;\;\; & & \mbox{} \nonumber \\
+g_{u_2 u_3}n_{u_1}n^{\ast}_{u_4}(n\tdot p_2 )
+g_{u_2 u_3}n_{u_4}n^{\ast}_{u_1}(n\tdot p_2 )\;\;\;\;\; & & \mbox{} \nonumber \\
-g_{u_2 u_3}n_{u_4}n_{u_1}(n^{\ast}\tdot p_2 )
-g_{u_2 u_3}n_{u_4}n_{u_1}(n^{\ast}\tdot p_3 )\;\;\;\;\; & & \mbox{} \nonumber \\
-g_{u_1 u_4}n_{u_3}n_{u_2}(n^{\ast}\tdot p_3 )
-g_{u_1 u_4}n_{u_3}n_{u_2}(n^{\ast}\tdot p_2 )\;\;\;\;\; & & \mbox{} \nonumber \\
+5\,(n_{u_3}n_{u_4}g_{u_1 u_2}(n^{\ast}\tdot p_3 )
+\,n_{u_1}n_{u_2}g_{u_3 u_4}(n^{\ast}\tdot p_2 )\;\;\;\;\; & & \mbox{} \nonumber \\
+\,n_{u_1}n_{u_2}g_{u_3 u_4}(n^{\ast}\tdot p_3 )
+\,n_{u_3}n_{u_4}g_{u_1 u_2}(n^{\ast}\tdot p_2 ))\;\;\;\;\; & & \mbox{} \nonumber \\
-4\,(n_{u_3}n_{u_1}g_{u_2 u_4}(n^{\ast}\tdot p_3 )
+\,n_{u_2}n_{u_4}g_{u_1 u_3}(n^{\ast}\tdot p_3 )\;\;\;\;\; & & \mbox{} \nonumber \\
+\,n_{u_2}n_{u_4}g_{u_1 u_3}(n^{\ast}\tdot p_2 )
+\,n_{u_3}n_{u_1}g_{u_2 u_4}(n^{\ast}\tdot p_2 ))\big]
\bigg{\}}.\;\;\;\;\; & & \mbox{} \nonumber 
\end{eqnarray}

\clearpage
%

%
\begin{ack}

We should like to acknowledge E. Yehudai for his software package
HIP-MAPLE, which was an invaluable tool for our work. One of us (C.P.M.)
would like to thank the Natural Sciences and Engineering Research
Council (NSERC) of Canada, and CICYT for financial support. The second
author gratefully acknowledges financial support in the form of a
scholarship from NSERC.  This research was supported in part by NSERC of
Canada under Grant No. A 8063.

\end{ack}
%

%

\end{document}